\documentclass[12pt, letterpaper, preprint]{aastex}
\usepackage[breaklinks,colorlinks, urlcolor=blue,citecolor=blue,linkcolor=blue]{hyperref}
\usepackage{color}
\usepackage{amsmath}
\usepackage{natbib}
\usepackage{ctable}
\usepackage{bm}
\usepackage[normalem]{ulem} 
\usepackage{xspace}

\let\oldbibliography\thebibliography 
\renewcommand{\thebibliography}[1]{%
  \oldbibliography{#1}%
  \setlength{\itemsep}{0pt}%
  \setlength{\parsep}{0pt}%
  \setlength{\parskip}{0pt}%
  \setlength{\bibsep}{0ex}
  \raggedright
}
\defcitealias{beutler2017}{B2017}
\defcitealias{sinha2017}{S2017}

\newcommand{\Nd}{{100}\xspace}

\newcommand{\specialcell}[2][c]{%
  \begin{tabular}[#1]{@{}c@{}}#2\end{tabular}}

\newcommand{\bitem}{\begin{itemize}}
\newcommand{\eitem}{\end{itemize}}
\newcommand{\beq}{\begin{equation}}
\newcommand{\eeq}{\end{equation}}
\newcommand{\Dmock}{\mathbf{D}^\mathrm{mock}}
\newcommand{\Xmock}{{\bf X}^\mathrm{mock}}
\newcommand{\Xref}{{\bf X}^\mathrm{ref}}
\newcommand{\Yref}{{\bf Y}^\mathrm{ref}}
\newcommand{\Ralpha}{R\'enyi-$\alpha$}

\newcommand{\Beut}{\citetalias{beutler2017}\xspace}
\newcommand{\Sinh}{\citetalias{sinha2017}\xspace}

\definecolor{orange}{rgb}{1,0.5,0}
\let\oldmarginpar\marginpar
\renewcommand\marginpar[1]{\-\oldmarginpar[\raggedleft\footnotesize #1]%
  {\raggedright\footnotesize #1}}

\newcommand{\lss}{{\small{LSS}}\xspace}
\newcommand{\gmm}{{\small{GMM}}\xspace}
\newcommand{\gmms}{{\small{GMM}s}\xspace}
\newcommand{\EM}{{\small{EM}}\xspace}
\newcommand{\bic}{{\small{BIC}}\xspace}
\newcommand{\pca}{{\small{PCA}}\xspace}
\newcommand{\ica}{{\small{ICA}}\xspace}

\newcommand{\patchy}{{\fontshape\scdefault\selectfont patchy}}

\begin{document}\sloppy\sloppypar\frenchspacing 

\title{Likelihood Non-Gaussianity in Large-Scale Structure Analyses}

\newcounter{affilcounter}

\setcounter{affilcounter}{1}

\edef \lbl {\arabic{affilcounter}}\stepcounter{affilcounter}
\altaffiltext{\lbl}{Lawrence Berkeley National Laboratory, 1 Cyclotron Rd, Berkeley CA 94720, USA}

\edef \bccp {\arabic{affilcounter}}\stepcounter{affilcounter}
\altaffiltext{\bccp}{Berkeley Center for Cosmological Physics, University of California, Berkeley, CA 94720, USA}

\edef \icg {\arabic{affilcounter}}\stepcounter{affilcounter}
\altaffiltext{\icg}{Institute of Cosmology \& Gravitation, Dennis Sciama Building, University of Portsmouth, Portsmouth PO1 3FX, UK}

\edef \swin {\arabic{affilcounter}}\stepcounter{affilcounter}
\altaffiltext{\swin}{Centre for Astrophysics \& Supercomputing, Swinburne University of Technology, 1 Alfred St., Hawthorn, VIC 3122, Australia}

\edef \astrothreed {\arabic{affilcounter}}\stepcounter{affilcounter}
\altaffiltext{\swin}{ARC Centre of Excellence for All Sky Astrophysics in 3 Dimensions (ASTRO 3D)}

\edef \vand {\arabic{affilcounter}}\stepcounter{affilcounter}
\altaffiltext{\vand}{Department of Physics and Astronomy, Vanderbilt University, Nashville, TN 37235, USA}

\edef \carn {\arabic{affilcounter}}\stepcounter{affilcounter}
\altaffiltext{\carn}{Department of Physics, Carnegie Mellon University, 5000 Forbes Avenue, Pittsburgh, PA 15213, USA}

\edef \berk {\arabic{affilcounter}}\stepcounter{affilcounter}
\altaffiltext{\bccp}{Department of Physics, University of California, Berkeley, CA 94720, USA}

\edef \ccpp {\arabic{affilcounter}}\stepcounter{affilcounter}
\altaffiltext{\ccpp}{Center for Cosmology and Particle Physics, New York University, New York, NY 10003, USA} 

\edef \cca {\arabic{affilcounter}}\stepcounter{affilcounter}
\altaffiltext{\cca}{Flatiron Institute, 162 Fifth Avenue, New York, NY 10010, USA} 

\author{
    ChangHoon~Hahn\altaffilmark{\lbl, \bccp},
    Florian~Beutler\altaffilmark{\icg, \lbl},
    Manodeep~Sinha\altaffilmark{\swin, \astrothreed, \vand}, 
    Andreas~Berlind\altaffilmark{\vand}, 
    Shirley~Ho\altaffilmark{\lbl, \bccp, \berk, \carn}, 
    David~W.~Hogg\altaffilmark{\ccpp, \cca}
}
\email{changhoon.hahn@lbl.gov}

\begin{abstract}
    Standard present day large-scale structure (LSS) analyses make a major 
    assumption in their Bayesian parameter inference --- that the likelihood has 
    a Gaussian form. For summary statistics currently used in LSS, this assumption, 
    even if the underlying density field is Gaussian, cannot be correct in detail. 
    We investigate the impact of this assumption on two recent LSS analyses: 
    the \cite{beutler2017}~power spectrum multipole ($P_\ell$) analysis and 
    the \cite{sinha2017}~group multiplicity function ($\zeta$) analysis. Using
    non-parametric divergence estimators on mock catalogs originally constructed 
    for covariance matrix estimation, we identify significant non-Gaussianity in 
    both the $P_\ell$ and $\zeta$ likelihoods. We then use Gaussian mixture 
    density estimation and Independent Component Analysis on the same mocks to 
    construct likelihood estimates that approximate the true likelihood better 
    than the Gaussian \emph{pseudo}-likelihood. Using these likelihood estimates, 
    we accurately estimate the true posterior probability 
    distribution of the \Beut~and~\Sinh~parameters. Likelihood non-Gaussianity 
    shifts the $f\sigma_8$ constraint by $-0.44\sigma$, but otherwise, does not 
    significantly impact the overall parameter constraints of~\Beut. For the 
    $\zeta$ analysis, using the pseudo-likelihood significantly underestimates 
    the uncertainties and biases the constraints of \Sinh~halo occupation parameters. 
    For $\log\, M_1$ and $\alpha$, the posteriors are shifted by $+0.43\sigma$ and 
    $-0.51\sigma$ and broadened by $42\%$ and $66\%$, respectively. The divergence 
    and likelihood estimation methods we present provide a straightforward framework 
    for quantifying the impact of likelihood non-Gaussianity and deriving more 
    accurate parameter constraints. 
\end{abstract}


\section{Introduction}
Bayesian parameter inference provides the standard framework for 
deriving cosmological parameters from observation of large scale 
structure (\lss) studies. 
Using Bayes' rule, 
\beq
p( \theta\,|\,x) \propto p( x\,|\,\theta) \, p(\theta)
\eeq
the posterior probability distributions of cosmological parameters
can be derived from observed measurements such as the galaxy 
power spectrum. All that is required are the prior distribution of the 
parameters, $p(\theta)$, and the likelihood, $p( x\,|\,\theta)$ --- 
probability of the data (observation) given the theoretical model.
Priors are selected in analyses; so parameter inference ultimately 
reduces to evaluating the likelihood. Analyses can {\em only} yield 
unbiased constraints if the likelihood evaluation is correct. 

In present day \lss analyses, two major assumptions go into evaluating the 
likelihood. First, the likelihood is assumed to have a Gaussian functional 
form: 
\beq \label{eq:gausslike}
p(x\,|\,\theta) = \frac{1}{(2\pi)^{d/2} \sqrt{\mathrm{det}(C)}}\,\,\mathrm{exp}\bigg[\frac{1}{2}(x - m(\theta))^{t} C^{-1} (x - m(\theta))\bigg]
\eeq
where $d$ is the dimension of the data vector $x$, $m(\theta)$ is the
theoretical predictions given the model parameters $\theta$, and $C$ 
is the covariance matrix. Second, the covariance matrix used in evaluating the Gaussian pseudo-likelihood
is assumed to be independent of cosmology or the model parameters.
The covariance matrix is evaluated only at a selected fiducial cosmology with  
fiducial model parameters and is assumed to be fixed throughout  
the analysis. In principle, the covariance matrix depends on $\theta$, 
and the dependence has been shown to have a significant effect 
on parameter constraints~\citep[\emph{e.g.}][]{eifler2009, morrison2013, white2015}. 
In this paper, we focus on the first, Gaussian pseudo-likelihood assumption. 
Even when analyses use covariance matrices that account for non-Gaussian 
covariance~\citep[\emph{e.g.}][]{scoccimarro1999, hu2001, oconnell2016}, 
the likelihood is still assumed to have a Gaussian functional form
(Eq.~\ref{eq:gausslike}). They therefore still employ a Gaussian \emph{pseudo}-likelihood. 
We will test the assumption and quantify the impact of this Gaussian
\emph{pseudo}-likelihood assumption on cosmological parameter constraints. 

The motivation for the Gaussian pseudo-likelihood ultimately stems 
from the `Central Limit Theorem'. Take the power spectrum of the density 
field for example. On large scales, the density field is approximately a 
Gaussian random field and the power spectrum of a specific Fourier 
mode would follow a chi-squared distribution, \emph{not} a Gaussian. 
However, with sufficiently many independent modes contributing, the likelihood
of the power spectrum would \emph{approach} a Gaussian distribution by the Central 
Limit Theorem. In practice, we expect the Gaussian assumption 
to fail in low signal-to-noise regimes. The assumption is also further
invalidated by correlations among different modes caused by 
by finite survey volume, shot noise, and systematic effects.
The breakdown of Gaussianity is clearly illustrated in earlier 
surveys such as IRAS where limited survey volume and sparse sampling cause the 
probability distribution function of the galaxy power spectrum to deviate 
significantly from Gaussian~\citep[see Figure 9 in][]{scoccimarro2000}. 
\cite{hartlap2009} and \cite{sellentin2017} similiarly illustrate the breakdown
of the Gaussian likelihood assumption for the cosmic shear correlation function 
likelihood.

Even if the likelihood is Gaussian, \cite{sellentin2016} argue that 
since an estimate of the covariance matrix is used for the likelihood, 
for accurate parameter inference the true covariance matrix must be marginalized over. 
This marginalization leads to a likelihood that is no longer Gaussian, but rather a 
multivariate $t$-distribution. Fortunately, the Gaussian pseudo-likelihood 
assumption is not necessary for parameter inference. Outside of \lss, in CMB 
power-specturm analyses for instance, the Planck collaboration 
uses a hybrid likelihood, which only assumes a Gaussian pseudo-likelihood 
for $C_\ell$ on small scales~(\citealt{ade2014,aghanim2016}; see 
also~\citealt{efstathiou2004, efstathiou2006}). On large scales (low-$\ell$), 
the likelihood is instead computed directly in pixel-space and extensively 
validated. Testing for likelihood non-Gaussianity and non-Gaussian likelihoods 
in general are not currently part of standard practice in \lss studies. For 
more precise parameter constraints from LSS, however, analyses must go beyond the
Gaussian pseudo-likelihood.

In this paper we investigate the impact of the likelihood Gaussianity
assumption on the two recent \lss analyses of \cite{beutler2017} (hereafter~\Beut)
and \cite{sinha2017} (hereafter~\Sinh). \Beut~analyzes the power spectrum 
multipoles ($P_\ell$; monopole, quadrupole, and hexadecapole) to measure
redshift-space distortions along with the Alcock-Paczynski effect and baryon 
acoustic oscillation scale. Meanwhile \Sinh~analyses the group multiplicity 
function ($\zeta$) in order to constrain parameters of the halo model.
Using the~\Beut~and~\Sinh~analyses, we show in this paper that the assumption 
of likelihood Gaussianity in \lss is not necessary. We will also show that 
the mock catalogs used in standard \lss analyses for covariance matrix estimation 
can be used to quantify the non-Gaussianity. More importantly, we will directly
use the mocks to estimate the ``true'' non-Gaussian likelihood. 

We begin in Section~\ref{sec:mocks} by describing the mock catalogs that
we use throughout the paper, constructed originally for covariance 
matrix estimation in~\Beut~and~\Sinh. Next in Section~\ref{sec:div}, we present
non-parametric divergence estimators and quantify the non-Gaussianity of
the $P_\ell$ and $\zeta$ likelihoods using them. Then in Section~\ref{sec:likeest}, 
we introduce two methods for estimating the ``true'' likelihood using the 
mock catalogs. We then use the likelihood estimates to quantify the impact 
of likelihood non-Gaussianity on the posterior parameter constraints 
of~\Beut~and~\Sinh in Section~\ref{sec:parinf}. We discuss and conclude 
the paper in Section~\ref{sec:summary}.  

\section{Mock Catalogs} \label{sec:mocks}
Mock catalogs are indispensable for standard cosmological 
analyses of \lss studies. They are used for testing analysis 
pipelines~\citep[][]{beutler2017, grieb2017, tinkerinpreparation}, 
testing the effect of systematics~\citep{guo2012, vargas-magana2014, hahn2017c, pinol2017, ross2017}, 
and, most relevantly for this paper, estimating the covariance 
matrix~\citep[][]{parkinson2012, kazin2014, grieb2017, alam2017, beutler2017, sinha2017}. 
In fact, nearly all current state-of-the-art \lss analyses use
covariance matrices estimated from mocks to evaluate the likelihood.

While some argue for analytic estimates of the covariance 
matrix~\citep[e.g.][]{mohammed2017} or estimates directly from data
by subsampling~\citep[e.g.][]{norberg2009}, covariance matrices 
from mocks have a number of advantages. Mocks allow us 
to incorporate detailed systematic errors present in the 
data as well as variance beyond the survey volume. Even for analytic 
estimates, a large ensemble of mocks are crucial for validation~\citep[\emph{e.g.}][]{slepian2017}. 
Moreover, as we show later in this paper, mocks present an
additional advantage: they allow us to 
quantify the non-Gaussianity of the likelihood and more accurately 
estimate the true likelihood distribution. 

In this paper, we focus on two \lss analyses: the power spectrum 
multipole ($P_\ell$) analysis of \Beut~and group multiplicity 
function ($\zeta$) analysis of \Sinh. Throughout the paper we will 
make extensive use of the mock catalogs used in these analyses. 
In this section, we give a brief description of these mocks and how 
the observables used in the analysis --- $P_\ell(k)$ and $\zeta(N)$ ---
are calculated from them. Afterwards, we will describe how we compute 
the covariance matrix from the mocks and pre-process the mock observable
data.

\subsection{MultiDark-PATCHY~Mock Catalog} \label{sec:patchy}
\Beut~use the MultiDark-\patchy~mock catalogs from \cite{kitaura2016}  mocks
generated using the \patchy~code \citep{kitaura2014,kitaura2015}. 
These mocks rely on large-scale density fields generated using augmented
Lagrangian Perturbation Theory~\citep[ALPT;][]{kitaura2013} on a mesh,
which are then populated with galaxies based on a combined non-linear
deterministic and stochastic biases. The mocks from the \patchy~code 
are calibrated to reproduce the galaxy clustering in the 
high-fidelity BigMultiDark $N$-body simulation~\citep{rodriguez-torres2016, klypin2016}. 
Afterwards, stellar masses are assigned to galaxies using the 
{\fontshape\scdefault\selectfont hadron}~code~\citep{zhao2015}.
Finally, the {\fontshape\scdefault\selectfont sugar}~code~\citep{rodriguez-torres2016} 
combines different boxes, incorporates selection
effects and masks to produce mock light-cone galaxy catalogs. 
The statistics of the resulting mocks are then compared to 
observations and the process is iterated to reach desired 
accuracy. We refer readers to~\cite{kitaura2016} for further 
details. 

In total, \cite{kitaura2016} generated 12,228 mock light-cone 
galaxy catalogs for BOSS Data Release 12. In \Beut, 
they use 2045 and 2048 for the northern galactic cap (NGC) and 
southern galactic cap (SGC) of the LOWZ+CMASS combined sample. 
\Beut~excluded 3 mock realizations due to notable 
issues. These issues have since been addressed so in our analysis 
we use all 2048 mocks for both the NGC and SGC of the LOWZ+CMASS
combined sample. In \Beut, they conduct multiple analyses, 
some using only the power spectrum monopole and quadrupole 
and others using monopole, quadrupole, and hexadecapole. They 
also separately analyze three redshift bins: $0.2 < z < 0.5$, 
$0.4 < z < 0.6$, and $0.5 < z < 0.75$. In this paper, for simplicity, 
we focus on one of these analyses: the analysis of the power spectrum 
monopole, quadrupole, and hexadecapole for the $0.2 < z < 0.5$ bin.  

\subsection{\cite{sinha2017} Mocks} \label{sec:gmf} 
The simulations used in the \cite{sinha2017} analysis are from the
Large Suite of Dark Matter Simulations project~\citep[LasDamas;][]{mcbride2009}, 
which were designed to model galaxy samples from SDSS DR7. The 
initial conditions are generated with the {\fontshape\scdefault\selectfont 2LTPIC} code~\citep{scoccimarro1998, crocce2006}, 
and evolved using the $N$-body $\mathtt{GADGET}$-$2$ code~\citep{springel2005}.
Halos are identified from the dark matter distribution outputs using 
the $\mathtt{ntropy-fofsv}$ code~\citep{gardner2007}, which uses a 
friend-of-friends algorithm~\citep[FoF;][]{davis1985} with a linking length of $0.2$
times the mean inter-particle separation. 
\Sinh~uses two configurations of the LasDamas simulations for the 
SDSS DR7 samples with absolute magnitude limits $M_r < -19$ and $M_r < -21$.
The `Consuelo' simulation contains $1400^3$ dark matter particles with 
mass of $1.87 \times 10^9\,h^{-1} M_\odot$ in a cubic volume of 
$420\,h^{-1} Mpc$ per side evolved from $z_\mathrm{init} = 99$. 
The `Carmen' simulation contains $1120^3$ dark matter particles with mass 
of $4.938 \times 10^{10}\,h^{-1} M_\odot$ in a cubic volume of 
$1000\,h^{-1} Mpc$ per side evolved from $z_\mathrm{init} = 49$. 

The FoF halo catalogs are populated with galaxies using the 
`Halo Occupation Distribution' (HOD) framework. The 
number, positions, and velocities of galaxies are described statistically 
by an HOD model. \Sinh~adopts the `vanilla' HOD model of \cite{zheng2007}, 
where the mean number of central and satellite galaxies are described by 
the halo mass and five HOD parameters: $M_\mathrm{min}, 
\sigma_{\log\,M} , M_0, M_1,~\mathrm{and}~\alpha$. Lastly, once the 
simulation boxes are populated with galaxies, observational systematic 
effects are imposed. The peculiar velocities of galaxies are used to 
impose redshift-space distortions. Galaxies that lie outside the redshift
limits or sky footprint of the SDSS sample are removed. For further 
details regarding the mocks, we refer readers to \Sinh.

To calculate their covariance matrix, \Sinh~produced 200 
independent mock catalogs from 50 simulations using a single set of 
HOD model parameters. The methods we propose in this paper rely on a large
number of mocks to accurately sample high dimensional distributions. 
We utilize an additional $99$ sets of HOD parameters, sampled from
the MCMC chain in~\Sinh, with $200$ mocks each. Thus, we
have a total of $20,000$ mocks for our current work.
In this paper we focus on the GMF analysis of the SDSS DR7 $M_r < -19$ sample
of presented in \Sinh.

\subsection{Mock Observable ${\bf X}^\mathrm{mock}$ and Covariance Matrix $\mathbb{C}$} \label{sec:xmock}
To get from the mock catalogs described above to the covariance matrices
used in~\Beut~and~\Sinh, the observables were measured for each mock in 
the \emph{same} way as the observations. We briefly describe how $P_\ell(k)$ and $\zeta(N)$ 
and the corresponding covariance matrices are measured in \Beut~and \Sinh. We then 
describe how we pre-process the mock observables for the methods we describe 
in the next sections.

To measure the power spectrum multipoles of the BOSS DR12 galaxies
and the MutliDark-\patchy~mocks (Section~\ref{sec:patchy}), \Beut~uses a 
fast Fourier transform (FFT) based anisotropic power spectrum estimator 
based on \cite{bianchi2015} and \cite{scoccimarro2015}. This estimator 
estimates the  monopole, quadrupole, and hexadecapole 
($\ell = 0, 2, 4$) of the power spectrum using FFTs of the 
overdensity field multipoles for a given survey geometry. For further 
details on the estimator we refer readers to Section 3 of~\Beut. The
power spectrum is computed in bins of $\Delta k = 0.01\,h\,\mathrm{Mpc}^{-1}$
over the range $k = 0.01 - 0.15\, h\,\mathrm{Mpc}^{-1}$
for $\ell = 0,\,\mathrm{and}\,2$ and $k = 0.01 - 0.10\, h\,\mathrm{Mpc}^{-1}$.
for $\ell = 4$. 
From the $\vec{P}^{(n)} = \big[P^{(n)}_{0}(k), P^{(n)}_{2}(k), P^{(n)}_{4}(k) \big]$ 
of the MultiDark-\patchy~mocks, \Beut~computes the $(i, j)$ element of the 
covariance matrix of all multipoles as 
\beq
\mathbb{C}_{i,j} = \frac{1}{N_\mathrm{mock} - 1} \sum\limits_{n=1}^{N_\mathrm{mock}} \big[ \vec{P}^{(n)}_i - \overline{P}_i \big]
    \times \big[ \vec{P}^{(n)}_j - \overline{P}_j \big].
\eeq
$N_\mathrm{mock} = 2048$ is the number of mocks and $\overline{P}_i$ is the mean 
of the mock powerspectra: $\overline{P}_i = \frac{1}{N_\mathrm{mock}} \sum_{n=1}^{N_\mathrm{mock}} \vec{P}^{(n)}_i$.
Since $P_0$ and $P_2$ each have $14$ bins and $P_4$ has $9$ bins, 
$\mathbb{C}$ is a $37 \times 37$ matrix.
In this work, we compute the $P_\ell(k)$ using a similar FFT-based 
estimator of \cite{hand2017a} instead of the~\Beut~estimator. Our choice 
is purely based on computational convenience. A python implementation
of the \cite{hand2017a} estimator is publicly available in the $\mathtt{NBODYKIT}$ 
package\footnote{http://nbodykit.readthedocs.io/en/latest/index.html}~\citep{hand2017b}. 
We confirm that the resulting $P_\ell(k)s$ and covariance matrices from 
the \cite{hand2017a} and \Beut~estimators are consistent with one another. 

Next, the \Sinh~group multiplicity function analysis starts 
with the \cite{berlind2006} FoF algorithm to identify groups in the 
SDSS and mock data. \Sinh~adopts the \cite{berlind2006} linking lengths
in units of mean inter-galaxy separation: $b_\perp = 0.14$ and $b_\parallel = 0.75$.
In comoving lengths, the linking lengths for the SDSS DR7 $M_r < -19$ 
sample correspond to $(r_\perp, r_\parallel) = (0.57, 3.05)h^{-1}\,\mathrm{Mpc}$. 
Once both the SDSS galaxy and mock galaxy groups are identified, $\zeta(N)$
is derived by calculating the comoving number density of groups 
in bins of richness $N$ --- the number of galaxies in a galaxy group. 
For the $M_r < −19$ sample, \Sinh~uses eight $N$ bins: 
$(5 - 6), (7 - 9), (10 - 13), (14 - 19), (20 - 32), (33 - 52), (53 - 84), (85 - 220)$.
For further details on the GMF calculation, we refer readers to 
Section 4.2 of \Sinh.  
From the $\zeta^{(n)}(N)s$ of each mock, \Sinh~computes the $(i, j)$ element of
the covariance matrix as 
\beq \label{eq:gmf_cov} 
\mathbb{C}_{i,j} = \frac{1}{N_\mathrm{mock} -1} \sum\limits_{n=1}^{N_\mathrm{mock}} \big[\zeta^{(n)}(N_i) -  \bar{\zeta}(N_i)\big] \times
    \big[\zeta^{(n)}(N_j) -  \bar{\zeta}(N_j)\big].
\eeq
In \Sinh, they compute the covariance matrix using $200$ mocks
generated using a single fiducial set of HOD parameters. As we 
describe in Section~\ref{sec:gmf}, in this paper we use $20,000$ mocks from 
$100$ different sets of HOD parameters sampled from the MCMC chain. 
The GMF covariance matrix we use in this paper is computed with 
$N_\mathrm{mock} = 20,000$ mocks. 

For the rest of the paper, in order to discuss the two separate 
analyses of~\Beut~and~\Sinh~in a consistent manner, we define the matrix 
$\Dmock$ of the mock observables ($P_\ell$ and $\zeta$) as 
\beq
\Dmock = \Big\{{\bf D}^\mathrm{mock}_n \Big\} \quad\quad \mathrm{where}\,\,\Dmock_n
\begin{cases}
    \vec{P}^{(n)} &\quad\mathrm{for}\,\, \mathrm{B}2017,\\ 
    \zeta^{(n)} &\quad\mathrm{for}\,\, \mathrm{S}2017.
\end{cases} 
\eeq
$\Dmock$ has dimensions of $2048 \times 37$ and $20,000\times8$ 
for \Beut~and \Sinh~respectively. 

For the methods in Sections~\ref{sec:gmm}~and~\ref{sec:ica}, the mock 
observable data ($\Dmock$) need to be pre-processed. This pre-processing
involves two steps: mean-subtraction (centering) and whitening. For mean subtraction, 
the mean of the observable is subtracted from $\Dmock$. Then 
$\Dmock - \overline{\bf D}^\mathrm{mock}$ is whitened using a linear transformation 
to remove the Gaussian correlation between the bins of $\Dmock$: 
\beq
\Xmock = L\,(\Dmock - \bar{\bf D}^\mathrm{mock}). 
\eeq
This linear transformation is derived such that the covariance matrix of the whitened 
data, $\Xmock$, is the identity matrix $\mathbb{I}$. Such a whitening linear 
transformation can be derived in infinite ways. 
One way to derive the linear transformation is through the eigen-decomposition
of the covariance matrix~\citep[\emph{e.g.}][]{hartlap2009, sellentin2017}. We, alternatively, 
derive the linear transformation ${\bf L}$ using Cholesky decomposition of the 
inverse covariance matrix~\citep{Press:1992:NRC:148286}: 
$\mathbb{C}^{-1} = {\bf L}\,{\bf L}^T$. We have checked that different methods 
for whitening do not impact the results of the paper. With this pre-processed 
mock observable data, we proceed to quantifying 
the non-Gaussianity of the $P_\ell$ and $\zeta$ likelihoods in the next section.

\section{Quantifying the Likelihood non-Gaussianity} \label{sec:div}
The standard approach to parameter inference in \lss studies does not 
account for likelihood non-Gaussianity. 
However, we are not the first to investigate likelihood non-Gaussianity 
in \lss analyses. Nearly two decades ago, \cite{scoccimarro2000} examined 
the likelihood non-Gaussianity for the power spectrum and reduced bispectrum 
using mock catalogs of the IRAS redshift catalogs. More recently, 
\cite{hartlap2009} and \cite{sellentin2017} examined the non-Gaussianity 
of the cosmic shear correlation function likelihood using simulations of 
the Chandra Deep Field South and CFHTLenS, respectively. 

While these works present different methods for identifying 
likelihood non-Gaussianity, they do not present a concrete way of 
quantifying it. \cite{hartlap2009}, for instance, identifies the
non-Gaussianity of the cosmic shear likelihood by looking at the
statistical independence/dependence of principal components of the mock 
observable. In \cite{sellentin2017}, they use the Mean
Integrated Squared Error (MISE) as a distance metric between 
Gaussian random variables and the whitened mock observable 
data vector to characterize non-Gaussian correlations between elements 
of the data vector. These indirect measures of non-Gaussianity 
are challenging to interpret or apply more generally to \lss studies. 

A more direct approach can be taken to quantify the 
non-Gaussianity of the likelihood. We can calculate the divergence between 
the distribution of our observable, $p(x)$, and $q(x)$ a multivariate Gaussian described 
by the average of the mocks and the covariance matrix --- \emph{i.e.} 
the pseudo-likelihood. The following are two of the most commonly used 
divergences: the Kullback-Leibler (KL) divergence
\beq \label{eq:kl} 
D_{KL} ( p \parallel q ) = \int p(x)\,\log \frac{p(x)}{q(x)}\,{\rm d}x
\eeq
and the R\'enyi-$\alpha$ divergence
\beq \label{eq:renyi}
D_{R-\alpha} ( p \parallel q ) = \frac{1}{\alpha -1} \log \int p^\alpha(x)\, q^{1 -\alpha}(x)\,{\rm d}x. 
\eeq
In the limit as $\alpha$ approaches 1, the R\'enyi-$\alpha$ divergence is
equivalent to the KL divergence.

Of course, in our case, we do not know $p(x)$ --- \emph{i.e.} the probability 
distribution function of our observable. If we did, we would simply use that 
instead of bothering with the covariance matrix or this paper. We can, however, 
still estimate the divergence using nonparametric divergence estimators~\citep{wang2009, poczos2012, krishnamurthy2014}. 
These estimators allow us to estimate the divergence, $\widehat{D}(X_{1:n} \parallel Y_{1:m})$, directly from 
samples $X_{1:n} = \{ X_1, ... X_n \}$ and $Y_{1:m} = \{ Y_1, ... Y_m \}$ 
drawn from $p$ and $q$ respectively.
For instance, the estimator presented in \cite{poczos2012} allows us to estimate 
the kernel function of the R\'enyi-$\alpha$ divergence,
\beq \label{eq:d_alpha}
D_{\alpha} ( p \parallel q ) = \int p^\alpha(x) q^{1-\alpha}(x)\,{\rm d}x. 
\eeq
using $k^\mathrm{th}$ nearest neighbor density estimators. Let $\rho_k(x)$ 
denote the Euclidean distance of the $k^\mathrm{th}$ nearest neighbor 
of $x$ in the sample $X_{1:n}$ and $\nu_k(x)$ denote the Euclidean distance 
of the $k^\mathrm{th}$ nearest neighbor of $x$ in the sample $Y_{1:m}$. Then 
\beq \label{eq:d_alpha_est}
D_{\alpha}(p \parallel q) \approx \widehat{D}_{\alpha}(X_{1:n} \parallel Y_{1:m}) = \frac{B_{k,\alpha}}{n} \left(\frac{n-1}{m}\right)^{1-\alpha}
\sum\limits_{i=1}^{n} \left(\frac{\rho_k^{d}(X_i)}{\nu_k^{d}(X_i)} \right)^{1-\alpha},
\eeq
where $B_{k, \alpha} = \dfrac{(\Gamma(k))^2}{\Gamma(k-\alpha+1)\Gamma(k+\alpha-1)}$. 
\cite{poczos2012} proves that this estimator is asymptotically unbiased: 
\beq
\lim_{n, m \rightarrow \infty} \mathbb{E} \big[ \widehat{D}_{\alpha} (X_{1:n} \parallel Y_{1:m}) \big] = D_{\alpha} (p \parallel q).
\eeq
Plugging $\widehat{D}_{\alpha}(X_{1:n} \parallel Y_{1:m})$ into Eq.~\ref{eq:renyi},
we get an estimator for the R\'enyi-$\alpha$ divergence. \cite{wang2009} derives
a similar estimator for the KL divergence~(Eq.~\ref{eq:kl}). 
These divergence estimates have been applied to Support Distribution Machines 
and used in the machine learning and astronomical literature with great success
\citep[\emph{e.g.}][]{poczos2011, poczos2012, poczos2012a, xu2013, ntampaka2015, ntampaka2016, ravanbakhsh2017a}. 
For more details on the non-parametric divergence estimators, we refer readers to 
\cite{poczos2012} and \cite{krishnamurthy2014}.

With these estimators, we can now explicitly quantify the 
non-Gaussianity of the likelihood by computing the divergence 
between the likelihood distribution and the Gaussian 
pseudo-likelihood distribution, $\mathcal{L}^\mathrm{pseudo}$.
$\Xmock$ is in principle sampled from $p(x)$. Then with a refrence 
sample ${\bf Y}^\mathrm{ref}$ drawn from $\mathcal{L}^\mathrm{pseudo}$, 
we can use the estimators to compute 
$D(\, p(x) \parallel \mathcal{L}^\mathrm{pseudo}) \approx \widehat{D}(\Xmock \parallel {\bf Y}^\mathrm{ref})$. 
Similar to the experiments detailed in \cite{poczos2012},
we construct ${\bf Y}^\mathrm{ref}$ with a comparable sample size as $\Xmock$:
$2000$ and $10,000$ for the $P_\ell$ and $\zeta$ analyses respectively.
For a sample size of $1000$, \cite{sutherland2012} use $k=5$. Based on the 
larger sample size of $\Xmock$, we calculate the divergences using the $k=10$ 
nearest neighbors. We note that the divergence estimates are not significantly 
impacted by our choice of $k$ within the range $5 < k < 20$.

\begin{figure}
\begin{center}
\includegraphics[width=0.9\textwidth]{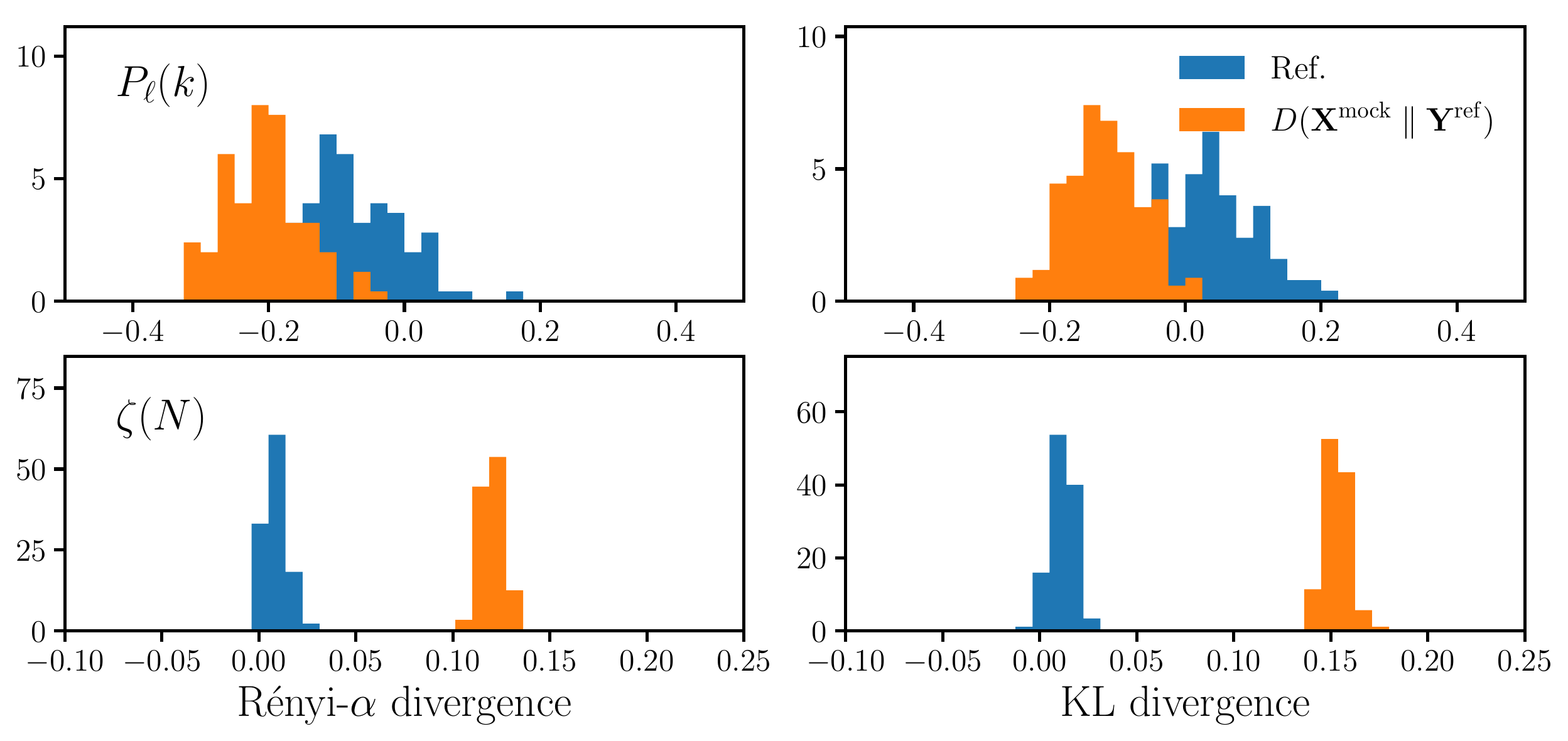}
    \caption{R\'enyi-$\alpha$ and KL divergence estimates 
    ($\widehat{D}_{R\alpha}$ and $\widehat{D}_{KL}$; orange) between 
    the likelihood distribution and the Gaussian pseudo-likelihood 
    for the \Beut~$P_\ell$ (top) and \Sinh~$\zeta$ (bottom) analyses. 
    We include in blue, as reference, the divergence estimates 
    of the pseudo-likelihood onto itself.  
    $\widehat{D}_{R\alpha}$ and $\widehat{D}_{KL}$ are computed using the 
    non-parametric $k$-NN estimator (Section~\ref{sec:div}) on 
    the mock data $\Xmock$ and a reference sample $\Yref$ drawn
    from the pseudo-likelihood. We compute $\widehat{D}_{R\alpha}$ and
    $\widehat{D}_{KL}$ \Nd times and plot their distribution 
    in order to illustrate the uncertainty of the $\widehat{D}$ 
    estimator. The significant discrepancy between 
    the two divergence distributions in each of the panels, identifies 
    the \emph{significant non-Gaussianity of the $P_\ell(k)$ and $\zeta(N)$ 
    likelihoods}. 
    }
\label{fig:div_gauss}
\end{center}
\end{figure}

In Figure~\ref{fig:div_gauss}, we present the resulting \Ralpha~(left) 
and KL (right) divergences (orange) between the likelihood and the 
Gaussian pseudo-likelihood for the \Beut~$P_\ell$ (top) and \Sinh~$\zeta$ (bottom) 
analyses: $\widehat{D}_{R\alpha}$ and $\widehat{D}_{KL}$. For reference, we also 
include (in blue) divergence estimates of the pseudo-likelihood onto itself, 
which we calculate as $\widehat{D}(\Xref \parallel \Yref)$. $\Xref$ is a 
data vector with the same dimension as $\Xmock$ sampled from the pseudo-likelihood. 
$\widehat{D}$s are \emph{estimates} of the true divergence, therefore we 
resample $\Yref$ and compute each $\widehat{D}$ estimate \Nd times. 
In Figure~\ref{fig:div_gauss}, we present the resulting distributions of $\widehat{D}$,
which illustrate the uncertainty of $\widehat{D}$. The discrepancy between 
the $\widehat{D}(\, p(x) \parallel \mathcal{L}^\mathrm{pseudo})$
distributions and the reference $\widehat{D}(\Xref \parallel \Yref)$ distributions 
($\Delta \widehat{D}$) quantify the discrepancy between the likelihood
and the pseudo-likelihood. Each panel of Figure~\ref{fig:div_gauss} shows significant 
discrepancy between the two distributions --- \emph{both 
the $P_\ell(k)$ and $\zeta(N)$ likelihoods are significantly non-Gaussian}. 

The Gaussian pseudo-likelihood assumption for $P_\ell$ is motivated 
by the Central Limit Theorem. If enough modes contribute to the 
power spectrum, then the likelihood approaches a Gaussian. Given the 
survey volume of BOSS DR12 and the restrictive $k$ range of the \Beut~analysis 
($0.01 < k < 0.15$ for $\ell = 0$ and $2$; $0.01 < k < 0.10$ for $\ell = 4$), 
one would expect this to be mostly true. Although relatively small, we
find significant $\Delta \widehat{D}$ and therefore likelihood non-Gaussianity. 
In order to better understand the source of this non-Gaussianity, 
we repeat the divergence comparisons for different $k$ ranges. If we 
exclude the largest scales and set $k_\mathrm{min} = 0.05$, $\Delta \widehat{D}$ 
decreases. Meanwhile, if we exclude the smallest scales and set 
$k_\mathrm{max} = 0.1$ for all 
multipoles, $\Delta \widehat{D}$ increases. This suggests that the largest 
scales (low $k$) contribute most to the $P_\ell$ likelihood non-Gaussianity.
Furthermore, when we compare the divergences for just the monopole and quadrupole, 
$\Delta \widehat{D}$ decreases. Among the multipole, the hexadecapole 
contributes most to the non-Gaussianity of the $P_\ell$ likelihood. 
In both the low $k$ regimes and the hexadecapole, the contribution to 
the non-Gaussianity is likely caused by low signal-to-noise and failure to 
satisfy the Central Limit Theorem.

For $\zeta$, the discrepancies between the $\widehat{D}$ distributions 
are consistent with the fact that the true $\zeta$ likelihood 
distribution is likely Poisson --- not Gaussian --- similar to the likelihood 
of observed cluster counts~\citep{cash1979,planckcollaboration2014,ade2016}. 
Although the groups identified with a FoF algorithm do not correspond
to clusters, we nevertheless expect the likelihood to be non-Gaussian. 
We again repeat the divergence comparison for different $N$ ranges 
to better understand the source of non-Gaussianity. Excluding the 
lowest $N$ bin does not significantly impact $\Delta \widehat{D}$. However, 
when we exclude the highest $N$ bin, $\Delta \widehat{D}$ decreases significantly. 
We therefore find that the high richness end of $\zeta$ contibute most to 
the non-Gaussianity of the $\zeta$ likelihood. The contribution to the 
non-Gaussianity, similar to the $P_\ell$ case, comes most from the low 
signal-to-noise regime. 
Besides likelihood non-Gaussianity, biases that arise from estimating
the covariance matrix from a limit number of mocks may also contribute 
to $\Delta \widehat{D}$. With $ > 2000$ mocks, however, this bias is 
likely unimportant for the $P_\ell$ analysis and even less so for the 
$\zeta$ analysis where we use $20,000$ mocks~\citep{hartlap2009}. Nonetheless, 
this underlines another limitation of using pseudo-likelihoods for 
parameter inference in \lss studies. 

\section{Estimating the Non-Gaussian Likelihood} \label{sec:likeest}
In the previous section, we estimate the divergence between the 
$P_\ell$ and $\zeta$ likelihoods and their respective Gaussian 
pseudo-likelihoods. These divergences identify and quantify 
the significant non-Gaussianity in the likelihoods of \lss studies.   
Our ultimate goals, however, are to quantify the impact of likelihood 
non-Gaussianity on the final cosmological parameter constraints and 
to develop more accurate methods for parameter inference in \lss. 
From the divergence estimates alone, it is not obvious how they propagate
onto the final parameter constraints. Therefore in this section,
we present two methods for more accurately estimating the true non-Gaussian
likelihoods of $P_\ell$ and $\zeta$ from their corresponding mocks.
These methods provide more accurate estmiates of the likelihood than
the Gaussian pseudo-likelihood. Moreover, we will use them later to
quantify the impact of likelihood non-Gaussianity on the
\Beut~and \Sinh~parameter constraints. 

\subsection{Gaussian Mixture Likelihood Estimation} \label{sec:gmm}
When mock catalogs are used for parameter inference in \lss analyses,
they essentially serve as data points sampling the likelihood distribution.
For the pseudo-likelihood, this distribution is assumed to have a
Gaussian functional form, which is why we estimate the covariance matrix 
from mocks. However, the Gaussian functional form, or any functional form for 
that matter, is \emph{not} necessary to estimate the likelihood distribution. 
Instead, the multi-dimensional likelihood distribution 
can be directly estimated from the set of mock catalogs --- for 
instance using Gaussian mixture density 
estimation~\citep{Press:1992:NRC:148286,9780471006268}. 
Besides its extensive use in machine learning and statistics, 
in astronomy, Gaussian mixture density estimation has been used for 
inferring the velocity distribution of stars from the Hipparcos 
satellite~\citep{bovy2011}, classifying galaxies in the Galaxy And Mass Assembly 
Survey~\citep{taylor2015}, classifying pulsars~\citep{lee2012}, and much more~\citep[see also][]{hogg2010,kuhn2017}. 

Gaussian mixture density estimation is a ``semi-parametric'' method 
that uses a weighted sum of $k$ Gaussian component densities, a Gaussian 
mixture model (hereafter \gmm)
\beq
\widehat{p}(x; \bm{\theta}) = \sum\limits_{i=1}^{k} \pi_i\, \mathcal{N}(x; \bm{\theta}_i),
\eeq
to estimate the density. 
The component weights ($\pi_i$; also known as mixing weights) and the 
component parameters $\bm{\theta}_i$ are free parameters of the mixture 
model. Given some data set ${\bf X}_N = \{{\bf x}_1,..., {\bf x}_N \}$, 
these free \gmm parameters are, most popularly, estimated 
through an expectation-maximization algorithm~\citep[\EM;][]{dempster1977, neal1998}.
The \EM algorithm begins by randomly assigning $\bm{\theta}^0_i$ to the 
$k$ Gaussian components. The algorithm then iterates between two steps. 
In the first step, the algorithm computes ${\bf x}_n$, 
a probability of being generated by each component of the model, for every data point. These 
probabilities can be thought of as weighted assignments of the points 
to the components. Next, given the ${\bf x}_n$ assignment to the 
components at some step $t$, $\bm{\theta}^t_i$ of each component are updated to $\bm{\theta}^{t+1}_i$
to maximize the likelihood of the assigned points. At this point, $\pi_i$ 
can also be updated by summing up the assignment weights and 
normalizing it by the total number of data points, $N$. This entire
process is repeated until convergence --- \emph{i.e.} when the log-likelihood of the
mixture model $\log\,p({\bf X}_N; \bm{\theta}^t)$ 
converges. The \EM algorithm is guaranteed to converges to a local maximum 
of the likelihood~\citep{wu1983}. In practice, instead of arbitrarily 
assigning the initial condition, $\bm{\theta}^0_i$ is derived from a $\texttt{k-means}$ 
clustering algorithm~\citep{lloyd1982}. The 
$\texttt{k-means}$ algorithm clusters a dataset, ${\bf X}_N$, into $k$ 
clusters, each described by the mean (or centroid) $\mu_i$ of the
samples in the cluster. The algorithm then iteratively chooses centroids that 
minimize the average squared distance between points in the same cluster.
For our \gmms, we initialize the \EM algorithm using 
the $\texttt{k-means++}$ algorithm of ~\cite{arthur2007}. 



So far in our description of \gmms, we have kept
the number of components $k$ fixed. $k$, however, is a free 
parameter and selecting $k$ is a crucial step in Gaussian mixture
density estimation. With too many components the model may overfit 
the data; while with too few components the model may not be flexible
enough to approximate the true 
underlying distribution. In order to address this model selection problem
when selecting $k$, we make use of the Bayesian Information 
Criterion~\citep[\bic;][]{schwarz1978}. \bic has been widely used for 
determining the number of components in mixture 
modeling~\citep[\emph{e.g.}][]{leroux1992,roeder1997,fraley1998,steele2010performance}
and for model selection in general in 
astronomy~\citep[\emph{e.g.}][]{liddle2007,broderick2011,wilkinson2015,vakili2016}.
According to \bic, models with higher likelihood are preferred; however, 
to address the concern of overfitting, \bic introduces a \emph{penalty} term 
for the number of parameters in the model: 
\beq \label{eq:bic}
\mathrm{BIC} = -2\,\mathrm{ln}\,\mathcal{L} + N_\mathrm{par}\,\mathrm{ln}\,N_\mathrm{data}.
\eeq
We select $k$ based on the number of components in the model with the 
lowest \bic. 


With Gaussian mixture density estimation we can directly estimate 
the likelihood distribution using the mock catalogs. We first fit 
\gmms with $k\mathrm{s} < 30$ components to the whitened 
mock data $\Xmock$ using the \EM algorithm for each model. For each of the 
converged \gmms, we calculate the \bic and then select the model with the 
lowest \bic as the best density estimate of the likelihood 
distribution: $\widehat{p}_\mathrm{\tiny GMM}(x)$. The selected density estimate can 
then be used to calculate the
likelihood and quantify the impact of likelihood non-Gaussianity on the 
parameter constraints of~\Beut~and~\Sinh. But first, we test whether  
$\widehat{p}_\mathrm{\tiny GMM}$ provides a better estimate of the non-Gaussian 
likelihoods over Gaussian pseudo-likelihoods by repeating the divergence estimates from 
Section~\ref{sec:div}.

\begin{figure}
\begin{center}
\includegraphics[width=0.9\textwidth]{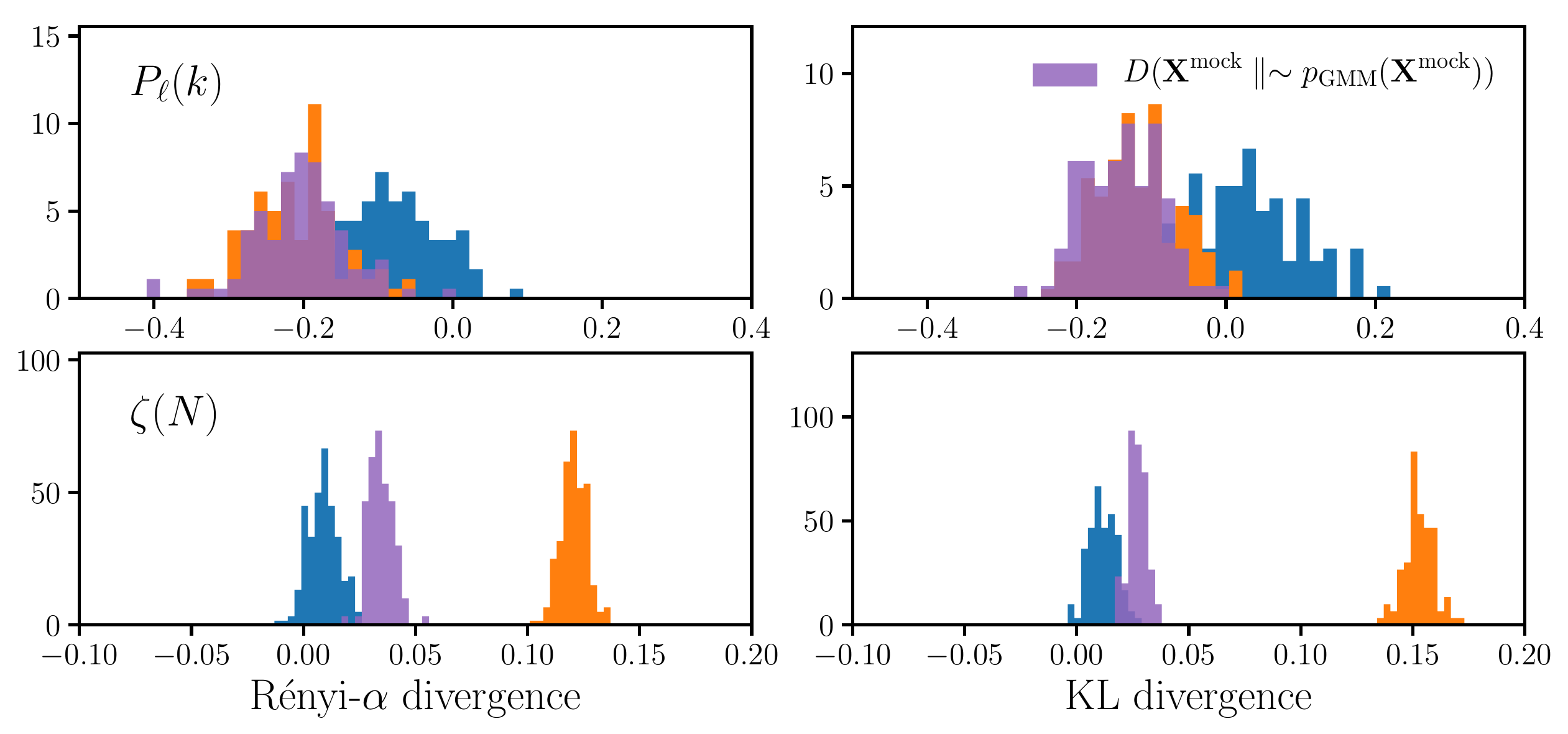}
\caption{R\'enyi-$\alpha$ and KL divergence estimates 
    ($\widehat{D}_{R\alpha}$ and $\widehat{D}_{KL}$; purple) between 
    the likelihood distribution and the Section~\ref{sec:gmm} GMM 
    likelihood estimate for the \Beut~$P_\ell$ (top) and \Sinh~$\zeta$ 
    (bottom) analyses. We include the divergence estimates for the 
    Gaussian pseudo-likelihood from 
    Figure~\ref{fig:div_gauss} (blue) for comparison. The Gaussian mixture
    likelihood does not significantly improve the discrepancy in 
    divergence for the $P_\ell$ analysis. This is due to the high-dimensionality 
    (37 dimensions) of the $P_\ell$ likelihood. For the $\zeta$ 
    analysis, \emph{our Gaussian mixture likelihood estimate is 
    a significantly better estimate of the likelihood
    than the pseudo-likelihood.} 
    }
\label{fig:div_gmm}
\end{center}
\end{figure}

To estimate the divergence between our Gaussian mixture density estimate,
$\widehat{p}_\mathrm{\tiny GMM}$, and the likelihood distribution, we take 
the same approach as our $\widehat{D}(\Xmock | \Yref)$ calculation in Section~\ref{sec:div}. 
Instead of $\Yref$ drawn from the pseudo-likelihood, we draw 
samples from $\widehat{p}_\mathrm{\tiny GMM}(x)$ with the same 
dimensions. Then we calculate $k$-NN~\Ralpha~and KL divergence estimates 
between this sample and $\Xmock$. To get a distribution of divergence
estimates that reflects the scatter in the estimator, we repeat the
estimates \Nd times resampling $\widehat{p}_\mathrm{\tiny GMM}$
each time (exactly the same method as for Figure~\ref{fig:div_gauss}). In Figure~\ref{fig:div_gmm}, 
we present the resulting distribution of divergences between 
$\widehat{p}_\mathrm{\tiny GMM}$ and the likelihood distribution in purple 
for the $P_\ell(k)$ (top) and $\zeta(N)$ (bottom) analyses. For comparison, 
we include the $\widehat{D}$ distributions for Gaussian pseudo-likelihoods from Figure~\ref{fig:div_gauss}.

From Figure~\ref{fig:div_gmm}, we see that the Gaussian mixture density 
estimate significantly improves the divergence discrepancy compared to the 
pseudo-likelihood for the $\zeta(N)$ analysis of \Sinh. In other words, 
\emph{our Gaussian mixture density estimate is a significant better estimate 
of the $\zeta$ likelihood distribution than the pseudo-likelihood}.
On the other hand, the Gaussian mixture 
density estimate for the $P_\ell(k)$ analysis of \Beut~does not 
significantly improve the divergence discrepancy. This difference in 
the performance of Gaussian mixture density estimation is not surprising. 
One would expect a direct density estimation to be more effective 
for the \Sinh~case, where we estimate an $8$-dimensional 
distribution with $N_\mathrm{mock} = 20,000$ samples, compared to the \Beut~ 
case where we estimate a $37$-dimensional distribution with 
only $N_\mathrm{mock} = 2048$ samples.
Given the unconvincing accuracy of the Gaussian mixture density estimate
of the $P_\ell$ likelihood, in the next section we present an alterative 
method for estimating the non-Gaussian likelihood.

\subsection{Independent Component Analysis} \label{sec:ica}
Gaussian mixture density estimation fails to accurately estimate 
the $37$-dimensional $P_\ell$ likelihood distribution of \Beut. 
Rather than estimating the likelihood distribution directly, 
if we can transform the observable ${\bf x}$ (\emph{e.g.} $P_\ell$) 
into statistically independent components ${\bf x}^\mathrm{IC}$ 
the problem becomes considerably simpler. Since ${\bf x}^\mathrm{IC}$ 
is statistically independent, the likelihood distribution becomes 
\beq \label{eq:ica_like}
p(x) = \prod\limits_{n=1}^{N_\mathrm{bin}} p_{x^\mathrm{IC}_n} (x) 
\eeq
where $N_\mathrm{bin}$ is the number of bins in the observable 
and the number of independent components. 
For the \Beut~case, this reduces the problem of estimating a 37 
dimensional distribution with $2048$ samples to a problem of 
estimating 37 one dimensional distributions with $2048$ samples 
each. The challenge is in {\em finding} such a transformation. 

Efforts in the past have attempted to tackle this sort of 
high-dimensional problem~\citep[\emph{e.g.}][]{scoccimarro2000,eisenstein2001,gaztanaga2005,norberg2009,sinha2017}.
They typically use singular value decomposition or principal 
component analysis~\citep[PCA;][]{Press:1992:NRC:148286}. For a Gaussian
likelihood, the PCA components of it are statistically independent. 
However, when the likelihood is \emph{not} Gaussian, the PCA components 
are uncorrelated but~\emph{not necessarily statistically independent}~\citep{hartlap2009}. 
Since the $P_\ell$ and $\zeta$ likelihoods are non-Gaussian, we cannot 
use \pca. Instead, we follow \cite{hartlap2009} and use Independent 
Component Analysis~\citep[\ica][]{herault1984,comon1994,hyvarinen2000,
hyvarinen2001independent}. 

In order to find the transformation of ${\bf x}$ to ${\bf x}^\mathrm{IC}$ 
we first assume that ${\bf x}$ is generated by some linear transformation
${\bf x} = {\bf M}\,{\bf x}^\mathrm{IC}$. Then the goal of \ica is to invert 
this problem, ${\bf y} = {\bf W}\,{\bf x}$, and find ${\bf W}$ and ${\bf y}$ 
that best estimate ${\bf x}^\mathrm{IC} \approx {\bf y}$. The basic 
premise of ICA is simple, \emph{maximizing non-Gaussianity maximizes the 
statistical independence}. Consider a single component of ${\bf y}$: 
\beq
{\bf y}_n = {\bf w}_n^{\tiny t}\,{\bf x} = {\bf w}_n^{\tiny t}\,{\bf M}\,{\bf x}^\mathrm{IC} 
\eeq
where ${\bf w}_n^{\tiny t}$ is the $n^\mathrm{th}$ row of ${\bf W}$. 
Since ${\bf y}_n$ is a linear combination of the independent 
components ${\bf x}^\mathrm{IC}$, from the Central Limit Theorem 
${\bf y}_n$ is necessarily more Gaussian than any of the 
components \emph{unless} ${\bf y}_n$ is equal to one of the 
${\bf x}^\mathrm{IC}$ components. In other words, we can achieve 
${\bf x}^\mathrm{IC} \approx {\bf y}$ by finding ${\bf W}$ that 
maximizes the non-Gaussianity of ${\bf y}$. For a more rigorous 
justification of ICA we refer readers to~\cite{hyvarinen2001independent}. 
In practice, non-Gaussianity is commonly measured using differential
entropy --- ``negentropy''. For ${\bf y}_n$ with density function 
$p_{y_n}$ the entropy is defined as
\beq
H_{y_n} =  - \int p_{y_n} (y)\, \log p_{y_n}(y)\, \mathrm{d}y. 
\eeq
Since the Gaussian distribution has the largest entropy among all 
distributions with a given variance, the negentropy can be defined 
as, 
\beq
J_{y_n} = H_{y_n^\mathrm{Gauss}} - H_{y_n}. 
\eeq
Finding the statistically independent components is now a matter
of finding the ${\bf W}$ that maximizes $\sum\limits_{n} J_{\bf y_n}$
--- the negentropy of ${\bf y}$. In this paper, we make use of the 
$\mathtt{FastICA}$ fixed-point iteration algorithm~\citep{hyvarinen1999}. 
The algorithm starts with randomly selected ${\bf w}_n$, then it uses 
approximations of negentropy from~\cite{hyvarinen1998} and Newton's method 
to iteratively solve for ${\bf W}$ that maximizes negentropy. For details 
on the $\mathtt{FastICA}$ algorithm, we refer readers to~\cite{hyvarinen1999}.

Performing ICA on the whitened observable data $\Xmock$, we derive the 
matrix ${\bf W}$ that transforms $\Xmock$ into $N_\mathrm{bin}$
approximately independent components: 
\beq
{\bf X}^\mathrm{ICA} = {\bf W}\,\Xmock = \{{\bf X}^\mathrm{ICA}_1, ..., {\bf X}^\mathrm{ICA}_{N_\mathrm{bin}}\}.
\eeq
From these statistically independent components and Eq.~\ref{eq:ica_like}, 
we can estimate the likelihood distribution. $p_{x^\mathrm{IC}_n} (x)$, 
from Eq.~\ref{eq:ica_like}, is the 1-dimensional distribution 
function of the $n^\mathrm{th}$ ICA component. This distribution 
is sampled by ${\bf X}^\mathrm{ICA}_n$, the transformed mock data. 
That means ${\bf X}^\mathrm{ICA}_n$ can be used to estimate 
$p_{x^\mathrm{ICA}_n}$ using a method like kernel 
density estimation~\citep[KDE;][]{9780387848587,feigelson2012}. 
With KDE, the density estimate, $\widehat{p}_{x^\mathrm{ICA}_n}$, is constructed by 
smoothing the empirical distribution of the ICA component $x^\mathrm{ICA}_n$ 
using a smooth kernel: 
\beq
\widehat{p}_{x^\mathrm{ICA}_n}(x) = \frac{1}{b\,N_\mathrm{mock}} \sum\limits_{j=1}^{N_\mathrm{mock}} K \left( \frac{x - \mathrm{X}^{(j),\mathrm{ICA}}_n}{b} \right). 
\eeq
$b$ is the bandwidth and $K$ is the kernel function. Following the 
choices of \cite{hartlap2009}, we use a Gaussian distribution for $K$ and the 
``rule of thumb'' bandwidth~\cite[also known as Scott's rule;][]{scott1992,davison2008} 
for $b$. Combining the $\widehat{p}_{x^\mathrm{ICA}_n}$ estimates for 
all $n = 1, ..., N_\mathrm{bin}$ into Eq.~\ref{eq:ica_like}, we 
can estimate the likelihood distribution $p(x) \approx \prod\limits_n \widehat{p}_{x^\mathrm{ICA}_n}(x)$

\begin{figure}
\begin{center}
\includegraphics[width=0.9\textwidth]{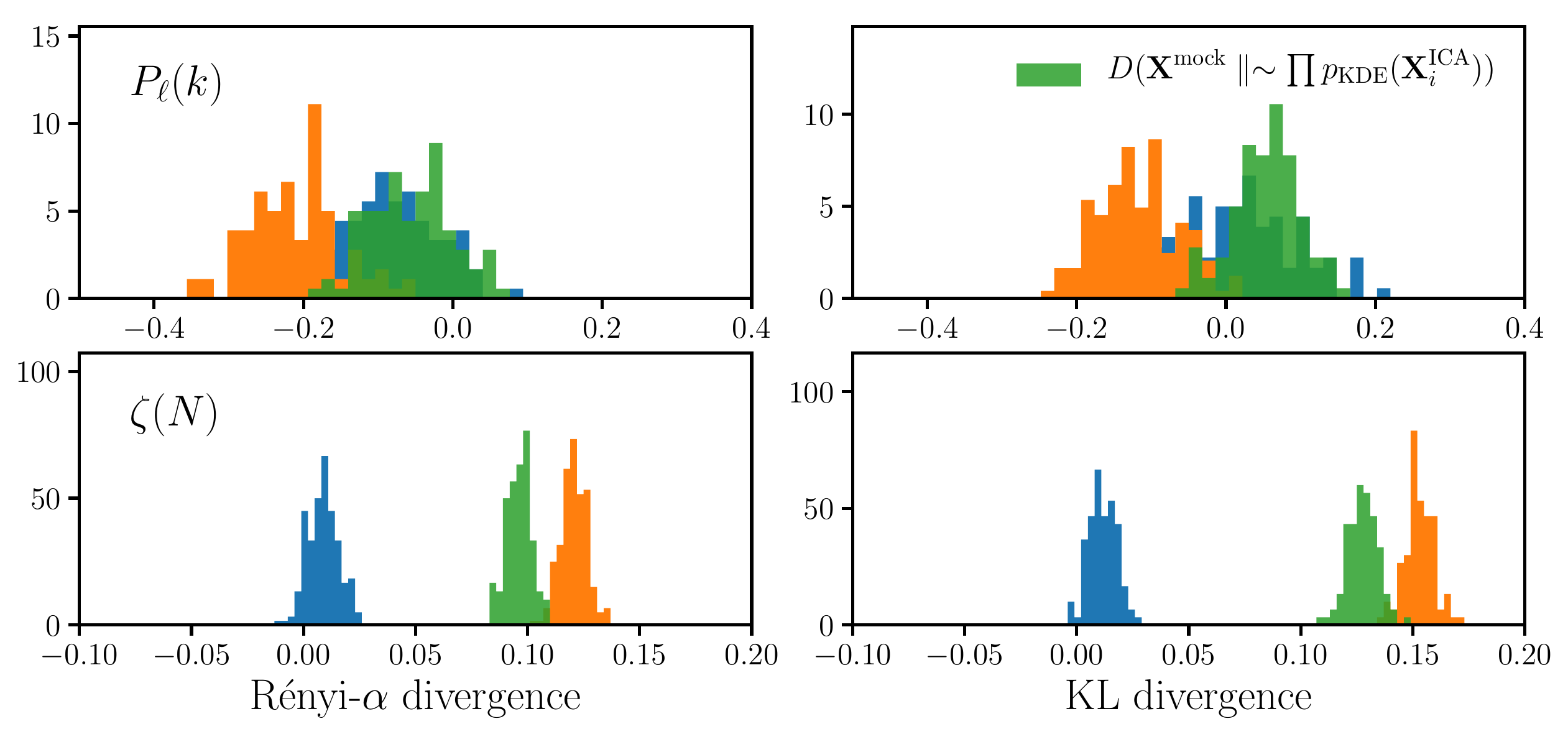}
\caption{R\'enyi-$\alpha$ and KL divergence estimates 
    ($\widehat{D}_{R\alpha}$ and $\widehat{D}_{KL}$; green) between 
    the likelihood distribution and the Section~\ref{sec:ica} \ica
    likelihood estimate for the \Beut~$P_\ell$ (top) and \Sinh~$\zeta$ 
    (bottom) analyses. We include the divergence estimates from 
    Figure~\ref{fig:div_gauss} for comparison. The \ica likelihood 
    significantly improves the divergence discrepancy for both the
    $P_\ell$ and $\zeta$ analyses. For $\zeta$, the improvement of 
    the \ica likelihood over the pseudo-likelihood is more modest than 
    our \gmm estimate from Section~\ref{sec:gmm}. However, for 
    $P_\ell$ where the GMM method struggled, our ICA likelihood 
    provides a significantly better estimate of the true $P_\ell$ 
    likelihood than the pseudo-likelihood.
    }
\label{fig:div_ica}
\end{center}
\end{figure}

We again check whether the likelihood estimate from \ica is actually 
a better estimate of the true likelihood distribution compared to the Gaussian 
pseudo-likelihood. Following the same procedure as we 
did for the Gaussian mixture likelihood in Section~\ref{sec:gmm}, we 
calculate the divergence between our ICA likelihood, $\prod \widehat{p}_{x^\mathrm{ICA}_n}(x)$, 
and the likelihood distribution, $p(x)$. We draw a sample from 
$\prod \widehat{p}_{x^\mathrm{ICA}_n}$ with the same dimensions as 
$\Yref$ (Section~\ref{sec:div}), apply the mixing matrix 
(undoing the \ica transformation), and then calculate the 
$k$-NN~\Ralpha~and KL divergence estimates between the sample and $\Xmock$. 
We repeat these steps \Nd times to get the distribution of estimates 
that reflects the scatter in the estimator. In Figure~\ref{fig:div_ica}, 
we present the resulting distribution of 
$\widehat{D}\left(\Xmock \parallel \sim \prod \widehat{p}_{x^\mathrm{ICA}_n} \right)$
in green for the $P_\ell(k)$ (top) and $\zeta(N)$ (bottom) analyses. 
For comparison, we include the distributions for the Gaussian pseudo-likelihood
from Figure~\ref{fig:div_gauss}. 

For \emph{both} \Beut~and~\Sinh, our \ica likelihood significantly 
improves the divergence discrepancy compared to the pseudo-likelihood. For 
\Sinh, however, the \ica likelihood proves to be less accurate than 
the Gaussian mixture likelihood in Section~\ref{sec:gmm}. More importantly, 
for \Beut where the Gaussian mixture likelihood did not improve 
upon the pseudo-likelihood, the \ica method provides a significantly 
more accurate likelihood estimate. This demonstrates that the \ica 
method is an effective alterative to the more direct Gaussian mixture method. 
The effectiveness of the \ica method in estimating
higher dimensional likelihoods with fewer samples (mocks) is  
particularly appealing for \lss, since analyses continue to 
increase the size of their observable data vector.
In~\cite{hartlap2009}, they suggest that a low $N_\mathrm{mock}$ may bias the 
ICA likelihood estimate. By examining the divergence discrpancy 
as we did in Figures~\ref{fig:div_ica} and~\ref{fig:div_gmm}, we ensure that 
pinpoint likelihood estimation methods that provide a better estimate of the true likelihood than the 
Gaussian pseudo-likelihood. Multiple methods can easily be tested to construct 
the best estimate of the likelihood distribution for {\em each }specific analysis. 
Based on the performances of the \gmm and \ica 
methods, we chose the \ica likelihood for the~\Beut~analysis 
and the \gmm likelihood for the~\Sinh~analysis. 

\section{Impact on Parameter Inference} \label{sec:parinf}
To derive the posterior distribution of their model parameters,
both~\Beut~and~\Sinh~use the standard Monte Carlo Markov Chain (MCMC) 
approach with the Gaussian pseudo-likelihood. The \Beut~analysis 
includes $11$ parameters,
\beq \nonumber
\Big \{f \sigma_8,~\alpha_\parallel,~\alpha_\perp,~b_1^\mathrm{NGC} \sigma_8,~b_1^\mathrm{SGC} \sigma_8, 
~b_2^\mathrm{NGC} \sigma_8,~b_2^\mathrm{SGC} \sigma_8,~\sigma_v^\mathrm{NGC},~\sigma_v^\mathrm{SGC}, 
~N^\mathrm{NGC},~\mathrm{and}~N^\mathrm{SGC} \Big \}, 
\eeq
while the \Sinh~analysis includes $5$ parameters,
\beq \nonumber
\Big\{ \log\,M_\mathrm{min},~\sigma_{\log\,M},~\log\,M_0,~\log\,M_1,~\mathrm{and}~\alpha \Big\}. 
\eeq
Using the improved likelihood estimates of Sections~\ref{sec:gmm} 
and~\ref{sec:ica}, we can now better estimate the true posteriors for the parameters
and quantify the impact of likelihood non-Gaussianity 
on parameter constraints. The ideal method to determine the true
posterior distributions would be to run new MCMC chains with non-Gaussian likelihood
estimators. While re-running MCMC chains is relatively tractable for the~\Beut~analysis, 
for~\Sinh~this is \emph{significantly} more involved. Rather than a perturbation 
theory based model from~\Beut, the \Sinh~model is a forward model, identical to
their mocks (Section~\ref{sec:gmf}). Re-running the MCMC samples would involve
evaluating the computationally costly forward model of~\Sinh~$\sim 10^6$ times
and is prohibitively expensive.

Without having to re-run the MCMC chains, we instead use importance sampling
to derive the new posteriors
from the original chains~\citep[see][for details on importance sampling]{wasserman2004}. 
The \emph{target} distribution we want is the new posterior. To sample this 
distribution, we re-weight the original posterior as the \emph{proposal} 
distribution with importance weights. In our case, the importance weights are the ratio 
of the (non-Gaussian) likelihood estimates over the (Gaussian) pseudo-likelihood. If we let 
$\mathrm{P}({\bf x} | \bm{\theta})$ be the original pseudo-likelihood and 
$\mathrm{P}'({\bf x} | \bm{\theta})$ be our ``new'' likelihood, then the new 
marginal likelihood can be calculated through importance sampling:   
\beq
\mathrm{P}'({\bf x} | \theta_1) = \int \mathrm{P}'({\bf x} | \bm{\theta})\,\mathrm{d}\theta_2...\mathrm{d}\theta_m = \int \frac{\mathrm{P}'({\bf x} | \bm{\theta})}{\mathrm{P}({\bf x} | \bm{\theta})}\, \mathrm{P}({\bf x} | \bm{\theta})\,\mathrm{d}\theta_2...\mathrm{d}\theta_m. \\
\eeq
Then through Monte Carlo integration, 
\beq
\mathrm{P}'({\bf x} | \theta_1) \approx \sum\limits_{\bm{\theta}^{(i)} \in S} \frac{\mathrm{P}'({\bf x} | \bm{\theta}^{(i)})}{\mathrm{P}({\bf x} | \bm{\theta}^{(i)})}. \label{eq:impsamp}
\eeq
where $S$ is the sample drawn from $\mathrm{P}({\bf x} | \bm{\theta})$. $S$ 
is simply the original MCMC chain in our case. The only calculation required 
is the importance weights in Eq.~\ref{eq:impsamp}, 
${\mathrm{P}'({\bf x} | \bm{\theta}^{(i)})}/{\mathrm{P}({\bf x} | \bm{\theta}^{(i)})}$ for 
each sample $\bm{\theta}^{(i)}$ of the original MCMC chain. 
For \Beut, $\mathrm{P}({\bf x} | \bm{\theta}^{(i)})$ is the \ica likelihood;  
for \Sinh, $\mathrm{P}({\bf x} | \bm{\theta}^{(i)})$ is the \gmm
likelihood.

\begin{figure}
\begin{center}
\includegraphics[width=\textwidth]{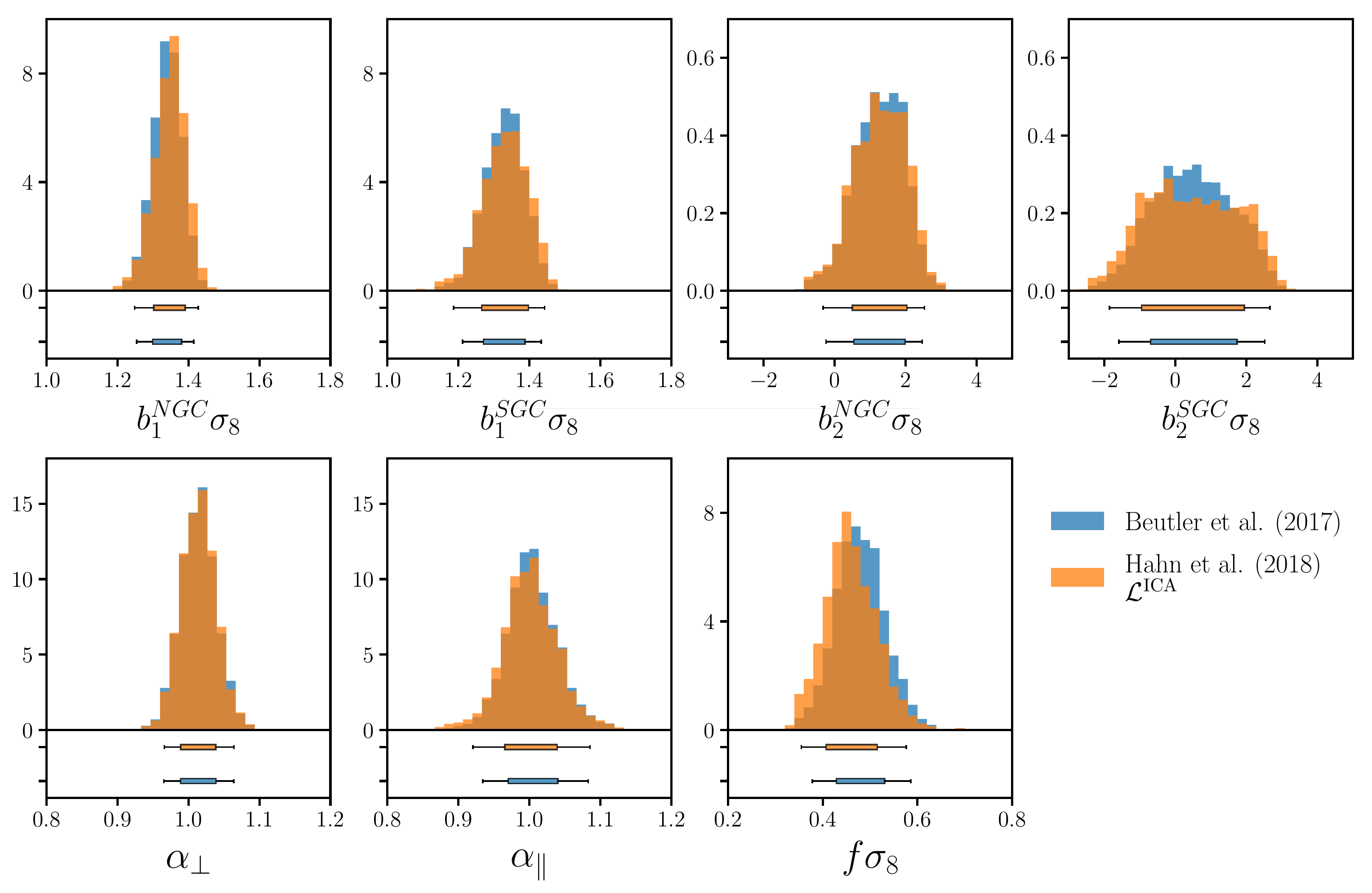}
\caption{The posterior distribution for 
    $\big \{f \sigma_8$, $\alpha_\parallel$, $\alpha_\perp$, 
    $b_1^\mathrm{NGC} \sigma_8$, $b_1^\mathrm{SGC} \sigma_8$, 
    $b_2^\mathrm{NGC} \sigma_8$, $b_2^\mathrm{SGC} \sigma_8,\big \}$ 
    in the~\Beut~$P_\ell$ analysis using the non-Gaussian \ica 
    likelihood (orange). We include in blue the original~\Beut~posteriors 
    for comparison. On the bottom of each panel we include box 
    plots that mark the $68\%$ and $95\%$ confidence intervals of
    the posterior. The discrepancies between the posteriors are most 
    evident for the parameters $f\sigma_8$ and and $b^\mathrm{SGC}_2 \sigma_8$.  
    The $f\sigma_8$ constraint shifts by $-0.44\sigma$. 
    Hence, using the pseudo-likelihood in the $P_\ell$ analysis 
    biases the posteriors of these parameters. 
    However, likelihood non-Gaussianity does \emph{not} 
    have a siginificant impact on the overall parameter constraints of 
    the $P_\ell$ analysis.} 
\label{fig:pk_like}
\end{center}
\end{figure}

In Figure~\ref{fig:pk_like} we present the resulting posterior 
distributions using the non-Gaussian \ica likelihood for the 
$\big \{f \sigma_8$, $\alpha_\parallel$, $\alpha_\perp$, 
$b_1^\mathrm{NGC} \sigma_8$, $b_1^\mathrm{SGC} \sigma_8$, 
$b_2^\mathrm{NGC} \sigma_8$, $b_2^\mathrm{SGC} \sigma_8\big \}$
parameters in the \Beut~$P_\ell$ analysis (orange). We include 
the original~\Beut~posteriors for comparison in blue. On the 
bottom of each panel, we also include box plots marking the 
confidence intervals of the updated and original posteriors. 
The boxes and ``whiskers'' repesent the $68\%$ and $95\%$ 
confidence intervals, respectively. The median and $68\%$ 
confidence intervals of the posteriors are also listed in Table~\ref{tab:posterior}.
$f \sigma_8$ and $b_2^\mathrm{SGC} \sigma_8$ are the main parameters with 
noticeable changes in their posteriors. After accounting for the 
non-Gaussian likelihood, the posterior of $b_2^\mathrm{SGC} \sigma_8$
broadens from $0.476^{+1.262}_{-1.175}$ to $0.422^{+1.517}_{-1.377}$.
More importantly, the $f \sigma_8$ posterior
shifts from $0.478^{+0.053}_{-0.049}$ to $0.456^{+0.059}_{-0.049}$, which 
corresponds to a shift of $-0.44 \sigma$. The other parameter constraints, 
however, remain largely unaffected by likelihood non-Gaussianity. 

\begin{figure}
\begin{center}
\includegraphics[width=\textwidth]{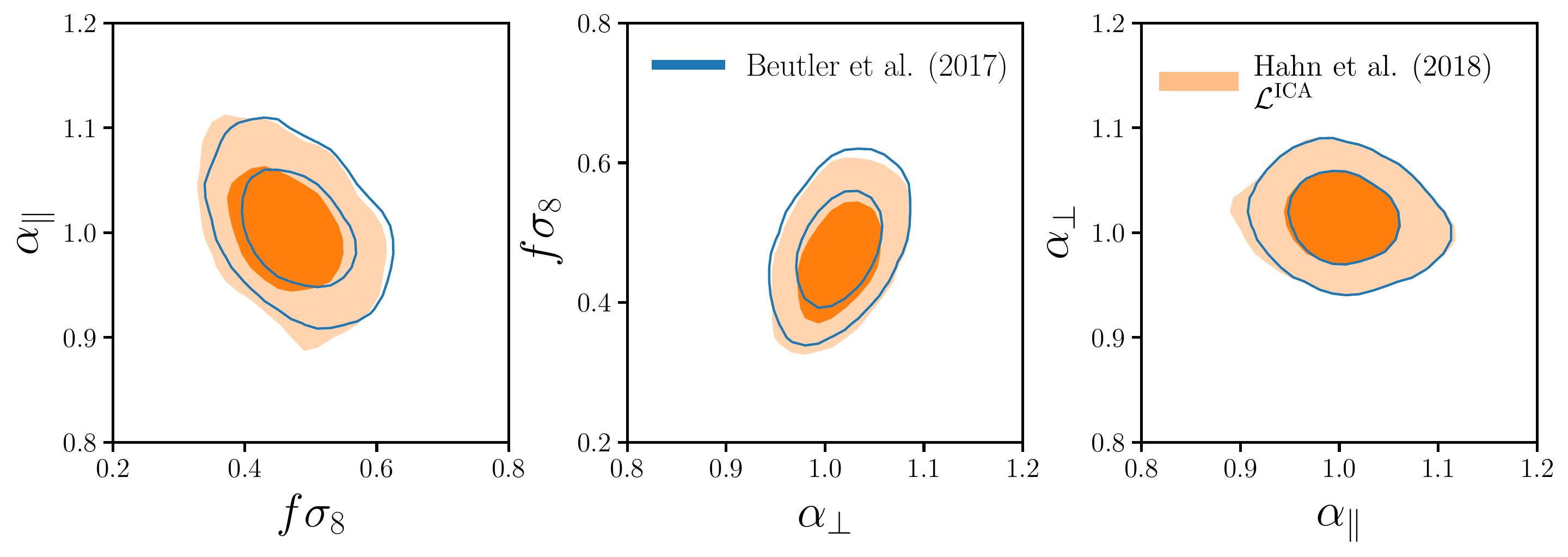}
\caption{Joint posterior distributions of $f \sigma_8$, 
    $\alpha_\parallel$, and $\alpha_\perp$ in the~\Beut~$P_\ell$ 
    analysis, compute using the non-Gaussian \ica likelihood (orange). 
    We include, in blue, the original~\Beut~posteriors for comparison. 
    The contours in the left and middle panels reflect the shift
    in $f \sigma_8$ caused by likelihood non-Gaussianity. Otherwise, 
    the contours illustrate that likelihood non-Gaussianity has little 
    impact on the cosmological parameters for the $P_\ell$ analysis.
    }
\label{fig:rsd_contour}
\end{center}
\end{figure}

Focusing on the main cosmological parameters $f \sigma_8$, 
$\alpha_\parallel$, and $\alpha_\perp$, we present their 
joint posterior distributions in Figure~\ref{fig:rsd_contour}.  
The contours mark the $68\%$ and $95\%$ confidence intervals
of the posteriors. The shift in the $f \sigma_8$ distribution 
is reflected in the
$(f\sigma_8, \alpha_\parallel)$ and $(\alpha_\perp, f\sigma_8)$ 
contours (left and middle panels respectively). The 
$(\alpha_\parallel, \alpha_\perp)$ distribution (right), however, show nearly 
no change from the non-Gaussian likelihood. 
Despite its impact on $f \sigma_8$ and $b_2^\mathrm{SGC} \sigma_8$, 
likelihood non-Gaussianity does \emph{not}
significantly impact the overall parameter constraints of the $P_\ell$ 
analysis. $b_2^\mathrm{SGC} \sigma_8$ is a poorly constrained nuisance 
parameter and although using the pseudo-likelihood biases $f \sigma_8$, 
the impact relative to its uncertainty is small --- less than $0.5 \sigma$. 
Furthermore, some of the impact may be from statistical fluctuation; although
this is likely not an important contributor since the \patchy~mocks are 
calibrated so that their $\overline{P_\ell}$ is consistent with the BOSS $P_\ell$. 
Some uncertainty is also introduced by the finite sampling of the MCMC chains. 
As mentioned in Section~\ref{sec:div}, some of the impact may also come 
from biases in covariance matrix estimation. Nevertheless, the fact that 
the $P_\ell$ analysis is largely unaffected by likelihood 
non-Gaussianity is consistent with the relatively small divergences found 
in Figure~\ref{fig:div_gauss}. It also illustrates
the remarkable effectiveness of the Central Limit Theorem. 

\begin{figure}
\begin{center}
\includegraphics[width=0.9\textwidth]{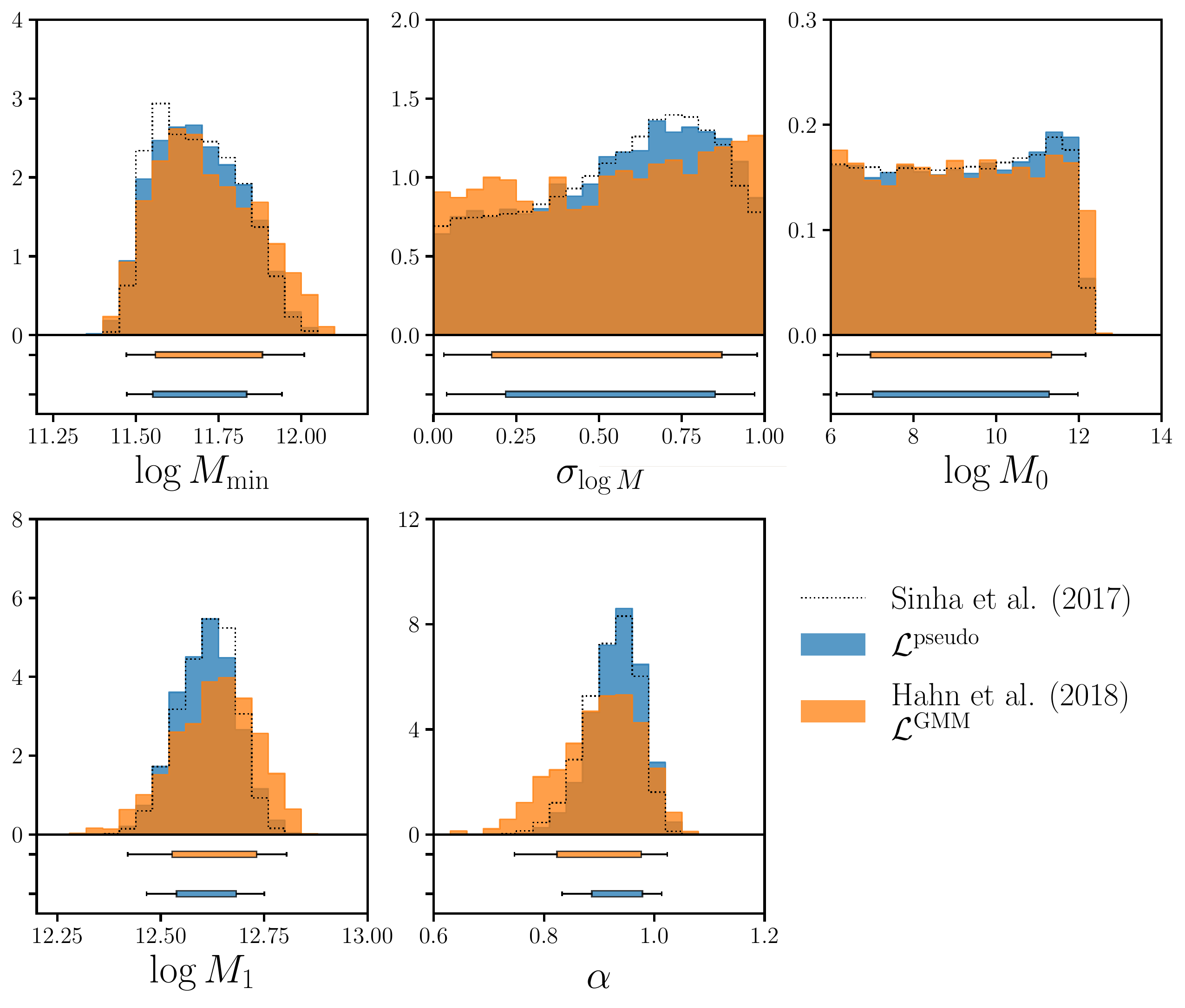}
\caption{The posterior distribution for HOD parameters 
    $\log\, M_\mathrm{min}$, $\sigma_{\log\,M}$, 
    $\log\, M_0$, $\log\, M_1$, and $\alpha$ in the~\Sinh~$\zeta$ 
    analysis using the non-Gaussian GMM likelihood (orange). We 
    include in blue the posteriors calculated from the pseudo-likelihood
    for comparison. We also include the original~\Sinh~posterior 
    (dotted; see text for details). On the bottom of each panel we 
    include box plots that mark the $68\%$ and $95\%$ confidence 
    intervals of the posterior. Besides the poorly constrained 
    parameters $\sigma_{\log M}$ and $\log\, M_0$, the posteriors of 
    $\log\,M_\mathrm{min}$, $\log M_1$, and $\alpha$, are significantly 
    broader and shifted compared to the pseudo-likelihood constraints. 
    Likelihood non-Gaussianity significantly impacts the parameter
    constraints of the $zeta$ analysis. Therefore using the pseudo-likelihood  
    underestimates the uncertainty and biases the HOD parameter constraints. 
    }
\label{fig:gmf_like}
\end{center}
\end{figure}

Next in Figure~\ref{fig:gmf_like}, we present the posterior 
distributions calculated using the non-Gaussian \gmm likelihood for 
the HOD parameters in the \Sinh~$\zeta$ analysis (orange).  
We include the posteriors calculated using the pseudo-likelihood 
for comparison in blue. The box plots on the bottom of each plot 
mark the $68\%$ and $95\%$ confidence intervals of the posteriors. 
In the dotted lines, we plot the original \Sinh~posteriors, 
which differ slightly from the blue distribution.
This discrepancy is caused by the difference in the covariance 
matrix we use in the pseudo-likelihood~(see Section~\ref{sec:xmock}).  
The difference, however, is negligible and goes to illustrate that the 
covariance matrix of $\zeta$ does not have a strong dependence on 
HOD parameters. In other words, our analysis is not significantly 
affected by our use of mocks generated from multiple HOD parameters. 

\begin{figure}
\begin{center}
\includegraphics[width=0.8\textwidth]{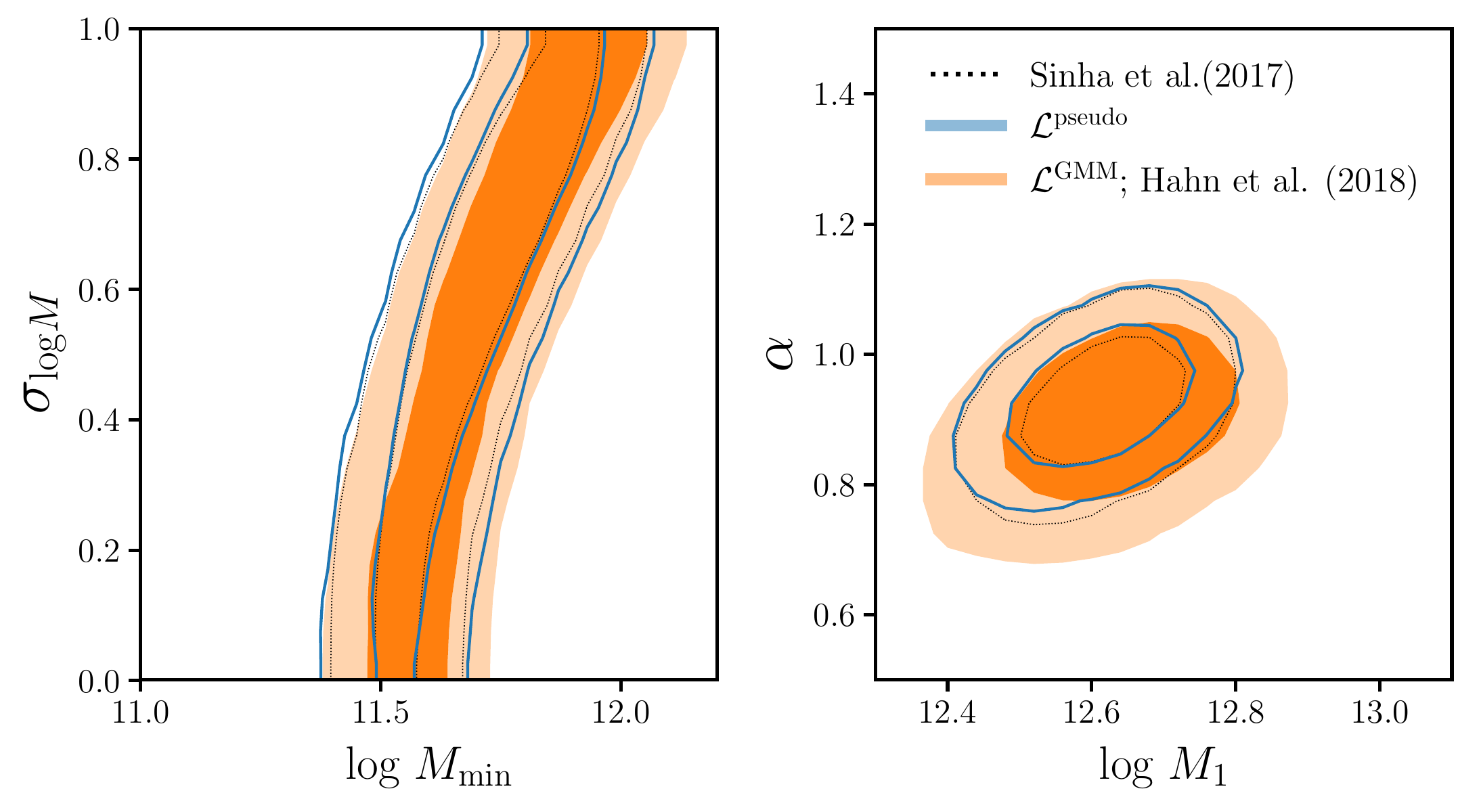}
\caption{Joint posterior distributions of select HOD parameters
    in the~\Sinh~$\zeta$ analysis, compute using the non-Gaussian 
    GMM likelihood (orange). We include, in blue, the posteriors 
    computed using the pseudo-likelihood; we also include the 
    original~\Sinh~posterior (dotted; see text for details). 
    The contours confirm that that \emph{due to likelihood 
    non-Gaussianity, posteriors from the pseudo-likelihood underestimate 
    the uncertainties and significantly biases the parameter constraints 
    of the \Sinh~analysis.} 
    }
\label{fig:gmf_contour}
\end{center}
\end{figure}

Besides the poorly constrained parameters $\sigma_{\log M}$ and $\log\, M_0$, 
likelihood non-Gaussianity significantly impacts the posterior 
distributions of the HOD parameters. Each of the parameter constraints 
for $\log\,M_\mathrm{min}$, $\log M_1$, and $\alpha$, are significantly 
broadened and shifted from the pseudo-likelihood constraints
(see Table~\ref{tab:posterior} for details). The $\log M_1$ constraint, 
for instance, is shifted by $+0.43 \sigma$ and its $68\%$ confidence
interval is expanded by $42\%$. Similarly, the $\alpha$ constraint is
shifted by $-0.51 \sigma$ and its $68\%$ confidence interval is expanded 
by $66\%$. The impact of likelihood non-Gaussianity is further emphasized in 
the joint posterior distributions in Figure~\ref{fig:gmf_contour}.
The $\log\,M_\mathrm{min}$ versus $\sigma_{\log M}$ and 
$\log\,M_\mathrm{min}$ versus $\alpha$ contours are both shifted 
and broadened compared to the $\mathcal{L}^\mathrm{pseudo}$ 
posterior. Figures~\ref{fig:gmf_like}~and~\ref{fig:gmf_contour}~reveal 
that \emph{using the Gaussian pseudo-likelihood significantly 
underestimates the uncertainty and biases the HOD parameter constraints 
of the \Sinh~$\zeta$ analysis.}

The contrast between the pseudo-likelihood posteriors and our posteriors
in Figures~\ref{fig:gmf_like}~and~\ref{fig:gmf_contour}~reflect 
the divergences in Figure~\ref{fig:div_gauss}, which revealed significant
discrepancy between the $\zeta$ likelihood and the pseudo-likelihood. 
These divergences and posteriors are consistent with the expectation 
that the true $\zeta$ likelihood distribution is likely Poisson. Although 
we expect the likelihood to be similar to the observed cluster count likelihood, 
the complicated connection between FoF groups and the underlying 
matter overdensity makes writing down the exact $\zeta$ likelihood function 
tremendously difficult. Nonetheless, the \gmm likelihood estimation method we 
present provides an accurate estimate of the non-Gaussian likelihood. 

The updated posteriors of the \Sinh~$\zeta$ analysis highlight the importance of 
accounting for likelihood non-Gaussianity in parameter inference of 
\lss studies. One of the main results of the \Sinh~HOD analysis 
is that the $\Lambda$CDM + HOD model can successfully 
fit either $\zeta(N)$ or the projected two-point correlation function 
$w_p(r_p)$ separately, but struggles to jointly fit both (see Figure~10 in \Sinh). 
Such a tension suggests that the `vanilla' HOD model is not sufficiently 
flexible in describing the galaxy-halo connection. Likelihood non-Gaussianity
is likely to impact this result. Once the non-Gaussianity is included in 
the analysis, the posteriors are broadened and shifted towards 
relaxing the tensions. We examine the effect of likelihood non-Gaussianity 
for HOD parameter constraints in more detail in Hahn et al. (in prep.). 

Even for the $P_\ell$ analysis, the impact of likelihood non-Gaussianity 
on the parameter constraints cannot be easily dismissed as we demand 
increasingly more precise constraints from future experiments. Using the 
pseudo-likelihood biases the $f\sigma_8$ constraints by $\sim 0.5\%$. 
Meanwhile, the Dark Energy Spectroscopic Instrument~\citep[DESI;][]{levi2013}, 
for instance, seeks to constrain $f\sigma_8$ to within a 
percent\footnote{DESI Final Design Report: http://desi.lbl.gov/wp-content/uploads/2014/04/fdr-science-biblatex.pdf}. 
The future, however, may be encouraging in this regard. The next surveys 
will expand the cosmic volumes probed by galaxies and therefore increase 
the number of modes on all scales. Even as they seek to extend the $k$ 
range of analyses, thanks to the Central Limit Theorm, we expect 
likelihood non-Gaussianity to have a smaller effect. However, without 
precisely quantifying the impact, as we have done in this paper, it 
remains to be determined whether likelihood non-Gaussianity will signficantly 
impact future $P_\ell$ analyses.  

For higher order statistics, likelihood non-Gaussianity will likely 
have a more significant effect. \cite{scoccimarro2000}   
found that the reduced bispectrum likelihood is significantly more 
non-Gaussian than the power spectrum likelihood. Constraints on primordial 
non-Gaussianity ($f_\mathrm{NL}$) from \lss~\citep[\emph{e.g.}][]{dalal2008, slosar2008, ross2013, giannantonio2014}, 
will also be significantly impacted by likelihood non-Gaussianity. 
In fact, the constraining power for $f_\mathrm{NL}$ comes from the 
largest scales -- the same scales that contribute most to the likelihood non-Gaussianity. 
Future experiments such as Euclid~\citep{amendola2016}, which seek 
to measure $\sigma(f_\mathrm{NL}) < 5$~\citep{giannantonio2012, amendola2016}, 
will need to robustly account for likelihood non-Gaussianity for 
accurate parameter constraints. Fortunately, the methods we present 
in this paper can easily be extended to other observables and analyses. 


\begin{table}
\caption{Impact of likelihood non-Gaussianity on the posterior parameter constraints of \Beut~and~\Sinh.}
\begin{center}
\begin{tabular}{cccccc} \toprule
    \multicolumn{6}{c}{\Beut~$P_\ell$ analysis} \\[5pt]
    \multicolumn{1}{c}{} & 
    \multicolumn{1}{c}{$b_1^\mathrm{NGC} \sigma_8$} & 
    \multicolumn{1}{c}{$b_1^\mathrm{SGC} \sigma_8$} &
    \multicolumn{1}{c}{$b_2^\mathrm{NGC} \sigma_8$} & 
    \multicolumn{1}{c}{$b_2^\mathrm{SGC} \sigma_8$} &
    \multicolumn{1}{c}{} \\
    \multicolumn{1}{c}{} & 
    \multicolumn{1}{c}{$f \sigma_8$} & 
    \multicolumn{1}{c}{$\alpha_\parallel$} & 
    \multicolumn{1}{c}{$\alpha_\perp$} &
    \multicolumn{1}{c}{} & 
    \multicolumn{1}{c}{} \\ [2pt]
    \hline \\[-10pt]
    \multicolumn{1}{c}{\Beut} & 
    \multicolumn{1}{c}{$1.341^{+0.040}_{-0.042}$} & 
    \multicolumn{1}{c}{$1.333^{+0.056}_{-0.062}$} &
    \multicolumn{1}{c}{$1.293^{+0.697}_{-0.752}$} & 
    \multicolumn{1}{c}{$0.476^{+1.262}_{-1.175}$} &
    \multicolumn{1}{c}{} \\ [3pt]
    \multicolumn{1}{c}{} & 
    \multicolumn{1}{c}{$0.478^{+0.053}_{-0.049}$} & 
    \multicolumn{1}{c}{$1.003^{+0.038}_{-0.032}$} & 
    \multicolumn{1}{c}{$1.014^{+0.025}_{-0.025}$} & 
    \multicolumn{1}{c}{} & 
    \multicolumn{1}{c}{} \\ [5pt]
    \multicolumn{1}{c}{\specialcell{non-Gaussian}} &  
    \multicolumn{1}{c}{$1.351^{+0.040}_{-0.049}$} & 
    \multicolumn{1}{c}{$1.335^{+0.063}_{-0.069}$} &
    \multicolumn{1}{c}{$1.295^{+0.746}_{-0.798}$} & 
    \multicolumn{1}{c}{$0.422^{+1.517}_{-1.377}$} &
    \multicolumn{1}{c}{} \\
    \multicolumn{1}{c}{$\mathcal{L}_\mathrm{ICA}$} & 
    \multicolumn{1}{c}{$0.456^{+0.059}_{-0.049}$} & 
    \multicolumn{1}{c}{$1.001^{+0.039}_{-0.035}$} & 
    \multicolumn{1}{c}{$1.014^{+0.024}_{-0.025}$} & 
    \multicolumn{1}{c}{} & 
    \multicolumn{1}{c}{} \\ [10pt]
    \hline 
    \hline \\[-10pt]
    \multicolumn{6}{c}{\Sinh~$\zeta$ analysis} \\
    \multicolumn{1}{c}{} & 
    \multicolumn{1}{c}{$\log\,M_\mathrm{min}$} & 
    \multicolumn{1}{c}{$\sigma_{\log\,M}$} &
    \multicolumn{1}{c}{$\log\,M_0$} & 
    \multicolumn{1}{c}{$\log\,M_1$} &
    \multicolumn{1}{c}{$\alpha$} \\ [2pt]
    \hline \\[-10pt]
    \multicolumn{1}{c}{\Sinh} & 
    \multicolumn{1}{c}{$11.68^{+0.148}_{-0.128}$} & 
    \multicolumn{1}{c}{$0.585^{+0.255}_{-0.367}$} & 
    \multicolumn{1}{c}{$9.154^{+2.074}_{-2.162}$} & 
    \multicolumn{1}{c}{$12.62^{+0.064}_{-0.077}$} & 
    \multicolumn{1}{c}{$0.928^{+0.042}_{-0.054}$} \\ [5pt]
    \multicolumn{1}{c}{\specialcell{Gaussian $\mathcal{L}_\mathrm{pseudo}$}} & 
    \multicolumn{1}{c}{$11.68^{+0.152}_{-0.131}$} & 
    \multicolumn{1}{c}{$0.586^{+0.264}_{-0.369}$} & 
    \multicolumn{1}{c}{$9.195^{+2.086}_{-2.180}$} & 
    \multicolumn{1}{c}{$12.61^{+0.070}_{-0.074}$} & 
    \multicolumn{1}{c}{$0.936^{+0.043}_{-0.049}$} \\ [5pt]
    \multicolumn{1}{c}{\specialcell{non-Gaussian \\[-2pt] $\mathcal{L}_\mathrm{GMM}$}} & 
    \multicolumn{1}{c}{$11.69^{+0.188}_{-0.135}$} & 
    \multicolumn{1}{c}{$0.554^{+0.317}_{-0.378}$} & 
    \multicolumn{1}{c}{$9.159^{+2.174}_{-2.198}$} & 
    \multicolumn{1}{c}{$12.64^{+0.095}_{-0.109}$} & 
    \multicolumn{1}{c}{$0.909^{+0.067}_{-0.086}$} \\ 
\hline
\end{tabular} \label{tab:posterior} 
\end{center}
\end{table}

\section{Summary and Discussion} \label{sec:summary}
Current \lss analyses makes a major assumption in their 
parameter inference --- the likelihood has a Gaussian functional-form. 
Although this assumption is motivated by the Central Limit Theorem, 
in detail the assumption cannot be true. 
In this paper, we investigate the impact of this Gaussian likelihood assumption on two recent 
\lss analyses: the \Beut~power spectrum multipole ($\ell = 0, 2$, and $4$) 
analysis and the \Sinh~group multiplicity function analysis. Using mock catalogs, 
originally constructed for covariance matrix estimation in these analyses, and 
non-parametric divergence estimators, used in Machine Learning, we measure the 
divergences between the $P_\ell$ and $\zeta$ likelihoods and the Gaussian 
pseudo-likelihoods from \Beut~and~\Sinh. For both the $P_\ell$ and $\zeta$ likelihoods, 
the divergences reveal significant likelihood non-Gaussianity. For the $P_\ell$ likelihood,
large scales (low $k$) and the hexadecapole contribute most to the
relatively small non-Gaussianity. For the $\zeta$ likelihood, 
the high richness end of $\zeta$ contribute most to the non-Gaussianity. 
In both likelihoods, we find that the low signal-to-noise regime contributes the most to the 
likelihood non-Gaussianity.

From the same mock catalogs of~\Beut~and~\Sinh, we estimate
the true non-Gaussian $P_\ell$ and $\zeta$ 
likelihoods with two different non-parametric density estimates -- Gaussian mixture density and
Independent Component Analysis. For the $\zeta$ likelihood, we find more
accurate estimates of the likelihood with the Gaussian mixture density method. 
For the \Beut~$P_\ell$ analysis, which has fewer mocks and a higher dimensional 
likelihood, we use Independent Component Analysis to transform the 
likelihood distribution into statistically independent components. 
By estimating the one-dimensional distribution of these independent 
components, we derive an estimate of the high-dimensional likelihood 
distribution for the \Beut~$P_\ell$ analysis. The divergence between our two likelihood estimates and 
the $P_\ell$ and $\zeta$ likelihoods demonstrate that we derive 
\emph{more accurate} estimates of the true likelihoods than the 
assumed Gaussian pseudo-likelihoods.

Finally, with these better estimates for the non-Gaussian $P_\ell$ and $\zeta$
likelihoods and importance sampling, we
calculate more accurate posterior parameter constraints for the \Beut~and~\Sinh 
analyses. 
By comparing our posteriors to the
parameter constraints from \Beut~and~\Sinh, we find that likelihood 
non-Gaussianity does not significantly impact the $P_\ell$ analysis of \Beut. 
Among the non-nuisance parameters, only $f\sigma_8$ is impacted by $-0.44\sigma$. 
Meanwhile for the \Sinh~$\zeta$ analysis, likelihood non-Gaussianity 
significant impacts the posterior distributions of the HOD parameter. 
Using the pseudo-likelihood significantly underestimates the width of the 
$\log\,M_\mathrm{min}$, $\log\, M_1$, and $\alpha$ posteriors and
significantly biases the \Sinh constraints. For $\log\, M_1$ and $\alpha$, the posteriors
are broadened by $42\%$ and $66\%$ and shifted by $+0.43\sigma$ and $-0.51\sigma$ 
respectively. Accounting for likelihood non-Gaussianity likely eases
the tension between the $\zeta$ and $w_p(r_p)$ constraints found 
in \Sinh. Our comparisons of the posteriors highlight the importance of 
incorporating likelihood non-Gaussianity in parameter inference of \lss studies. 

Based on our results, it is unclear whether $P_\ell$ analyses using future 
surveys (\emph{e.g.} DESI, Euclid) will be significantly impacted by 
likelihood non-Gaussianity. While future surveys will expand the cosmic
volumes probed by galaxies and increase the number of modes on all scales, 
future analyses also seek to extend the $k$ ranges probed and demand
unprecedentedly precise constraints. Meanwhile, for $\zeta$ analyses with
the same multiplicity range, we expect future surveys to reduce the impact of
likelihood non-Gaussianity, since larger cosmic volumes will probe more
high multiplicity groups. For a wider multiplicity range, however, likelihood
non-Gaussianity may still be a significant effect. For higher order statistics
such as the galaxy bispectrum or three-point function, even for future surveys,
likelihood non-Gaussianity will likely be a significant effect to consider for
parameter inference. We also expect it to significantly impact primordial non-Gaussianity
($f_\mathrm{NL}$) constraints from \lss, which derive most of their constraining
power from the largest, most non-Gaussian, scales. Regardless of our expectation,
for more accurate parameter inference the Gaussian likelihood assumption must
be extensively tested. The divergence and likelihood estimations we introduce in 
this paper a straightforward framework for testing and quantifying the 
impact of likelihood non-Gaussianity on the final parameter constraints. 

Our likelihood estimation methods also allow us to go beyond the pseudo-likelihood 
and derive more accurate estimates of the likelihood. With a similar 
motivation at addressing likelihoods that are non-Gaussian or difficult to 
write down, methods for likelihood-free inference such as Approximate Bayesian 
Computation~\citep[ABC;][]{hahn2017b,kacprzak2017,alsing2018} have recently 
been introduced to \lss studies. 
Although as a likelihood-free inference method, ABC has the advantage 
of relaxing \emph{any} assumption on the likelihood, even with smart sampling methods 
like Population Monte Carlo, it requires an expensive generative forward 
model to be computed far more times than the number of mocks required for 
covariance matrix estimation. Our methods (especially the \ica method)
do not require any more mocks than those already constructed for accurate covariance
matrix estimation. Therefore, the methods for likelihood estimation we present
in this paper provide both accurate and practical methods for Bayesian 
parameter inference in \lss. 

\section*{Acknowledgements}
It's a pleasure to thank 
    Emanuele~Castorina,
    Yu~Feng, 
    Simone~Ferraro,
    Daniel~Foreman-Mackey, 
    Emmanuel~Schaan, 
    Roman~Scoccimarro,
    Uro{\u s}~Seljak,
    Sukhdeep~Singh, 
    Michael~Wilson, 
    and Martin~White
for valuable discussions.
This material is based upon work supported by the U.S. Department 
of Energy, Office of Science, Office of High Energy Physics, under 
contract No. DE-AC02-05CH11231.
This project used resources of the National Energy Research 
Scientific Computing Center, a DOE Office of Science User 
Facility supported by the Office of Science of the U.S. 
Department of Energy under Contract No. DE-AC02-05CH11231.
Parts of this research were conducted by the Australian Research Council
Centre of Excellence for All Sky Astrophysics in 3 Dimensions (ASTRO 3D),
through project number CE170100013.
This project also made use of the NASA Astrophysics Data 
System and open-source software Python, numpy, SciPy, matplotlib, 
and Scikit-learn. 

\bibliographystyle{yahapj}
\bibliography{nongausslike}

\begin{thebibliography}{}
\providecommand\natexlab[1]{#1}
\providecommand\JournalTitle[1]{#1}

\bibitem[{Ade {et~al.}(2014)Ade, Aghanim, Armitage-Caplan, Arnaud, Ashdown,
  Atrio-Barandela, Aumont, Baccigalupi, Banday, Barreiro, Bartlett, Battaner,
  Benabed, Beno{\^\i}t, Benoit-L{\'e}vy, Bernard, Bersanelli, Bielewicz, Bobin,
  Bock, Bonaldi, Bonavera, Bond, Borrill, Bouchet, Boulanger, Bridges, Bucher,
  Burigana, Butler, Calabrese, Cardoso, Catalano, Challinor, Chamballu, Chiang,
  Chiang, Christensen, Church, Clements, Colombi, Colombo, Combet, Couchot,
  Coulais, Crill, Curto, Cuttaia, Danese, Davies, Davis, de~Bernardis, de~Rosa,
  de~Zotti, Delabrouille, Delouis, D{\'e}sert, Dickinson, Diego, Dole,
  Donzelli, Dor{\'e}, Douspis, Dunkley, Dupac, Efstathiou, Elsner, En{\ss}lin,
  Eriksen, Finelli, Forni, Frailis, Fraisse, Franceschi, Gaier, Galeotta,
  Galli, Ganga, Giard, Giardino, Giraud-H{\'e}raud, Gjerl{\o}w,
  Gonz{\'a}lez-Nuevo, G{\'o}rski, Gratton, Gregorio, Gruppuso, Gudmundsson,
  Hansen, Hanson, Harrison, Helou, Henrot-Versill{\'e},
  Hern{\'a}ndez-Monteagudo, Herranz, Hildebrandt, Hivon, Hobson, Holmes,
  Hornstrup, Hovest, Huffenberger, Hurier, Jaffe, Jaffe, Jewell, Jones, Juvela,
  Keih{\"a}nen, Keskitalo, Kiiveri, Kisner, Kneissl, Knoche, Knox, Kunz,
  Kurki-Suonio, Lagache, L{\"a}hteenm{\"a}ki, Lamarre, Lasenby, Lattanzi,
  Laureijs, Lawrence, Jeune, Leach, Leahy, Leonardi, Le{\'o}n-Tavares,
  Lesgourgues, Liguori, Lilje, Linden-V{\o}rnle, Lindholm, L{\'o}pez-Caniego,
  Lubin, Mac{\'\i}as-P{\'e}rez, Maffei, Maino, Mandolesi, Marinucci, Maris,
  Marshall, Martin, Mart{\'\i}nez-Gonz{\'a}lez, Masi, Massardi, Matarrese,
  Matthai, Mazzotta, Meinhold, Melchiorri, Mendes, Menegoni, Mennella,
  Migliaccio, Millea, Mitra, Miville-Desch{\^e}nes, Molinari, Moneti, Montier,
  Morgante, Mortlock, Moss, Munshi, Murphy, Naselsky, Nati, Natoli,
  Netterfield, N{\o}rgaard-Nielsen, Noviello, Novikov, Novikov, O'Dwyer,
  Orieux, Osborne, Oxborrow, Paci, Pagano, Pajot, Paladini, Paoletti,
  Partridge, Pasian, Patanchon, Paykari, Perdereau, Perotto, Perrotta,
  Piacentini, Piat, Pierpaoli, Pietrobon, Plaszczynski, Pointecouteau, Polenta,
  Ponthieu, Popa, Poutanen, Pratt, Pr{\'e}zeau, Prunet, Puget, Rachen, Rahlin,
  Rebolo, Reinecke, Remazeilles, Renault, Ricciardi, Riller, Ringeval,
  Ristorcelli, Rocha, Rosset, Roudier, Rowan-Robinson, Rubi{\~n}o-Mart{\'\i}n,
  Rusholme, Sandri, Sanselme, Santos, Savini, Scott, Seiffert, Shellard,
  Spencer, Starck, Stolyarov, Stompor, Sudiwala, Sureau, Sutton, Suur-Uski,
  Sygnet, Tauber, Tavagnacco, Terenzi, Toffolatti, Tomasi, Tristram, Tucci,
  Tuovinen, T{\"u}rler, Valenziano, Valiviita, Tent, Varis, Vielva, Villa,
  Vittorio, Wade, Wandelt, Wehus, White, White, Yvon, Zacchei, \&
  Zonca}]{ade2014}
Ade, P. a.~R., Aghanim, N., Armitage-Caplan, C., {et~al.} 2014,
  \href{http://dx.doi.org/10.1051/0004-6361/201321573}{\JournalTitle{Astronomy
  \& Astrophysics}, 571, A15}

\bibitem[{Ade {et~al.}(2016)Ade, Aghanim, Arnaud, Ashdown, Aumont, Baccigalupi,
  Banday, Barreiro, Bartlett, Bartolo, Battaner, Battye, Benabed, Beno{\^\i}t,
  Benoit-L{\'e}vy, Bernard, Bersanelli, Bielewicz, Bock, Bonaldi, Bonavera,
  Bond, Borrill, Bouchet, Bucher, Burigana, Butler, Calabrese, Cardoso,
  Catalano, Challinor, Chamballu, Chary, Chiang, Christensen, Church, Clements,
  Colombi, Colombo, Combet, Comis, Couchot, Coulais, Crill, Curto, Cuttaia,
  Danese, Davies, Davis, de~Bernardis, de~Rosa, de~Zotti, Delabrouille,
  D{\'e}sert, Diego, Dolag, Dole, Donzelli, Dor{\'e}, Douspis, Ducout, Dupac,
  Efstathiou, Elsner, En{\ss}lin, Eriksen, Falgarone, Fergusson, Finelli,
  Forni, Frailis, Fraisse, Franceschi, Frejsel, Galeotta, Galli, Ganga, Giard,
  Giraud-H{\'e}raud, Gjerl{\o}w, Gonz{\'a}lez-Nuevo, G{\'o}rski, Gratton,
  Gregorio, Gruppuso, Gudmundsson, Hansen, Hanson, Harrison,
  Henrot-Versill{\'e}, Hern{\'a}ndez-Monteagudo, Herranz, Hildebrandt, Hivon,
  Hobson, Holmes, Hornstrup, Hovest, Huffenberger, Hurier, Jaffe, Jaffe, Jones,
  Juvela, Keih{\"a}nen, Keskitalo, Kisner, Kneissl, Knoche, Kunz, Kurki-Suonio,
  Lagache, L{\"a}hteenm{\"a}ki, Lamarre, Lasenby, Lattanzi, Lawrence, Leonardi,
  Lesgourgues, Levrier, Liguori, Lilje, Linden-V{\o}rnle, L{\'o}pez-Caniego,
  Lubin, Mac{\'\i}as-P{\'e}rez, Maggio, Maino, Mandolesi, Mangilli, Maris,
  Martin, Mart{\'\i}nez-Gonz{\'a}lez, Masi, Matarrese, McGehee, Meinhold,
  Melchiorri, Melin, Mendes, Mennella, Migliaccio, Mitra,
  Miville-Desch{\^e}nes, Moneti, Montier, Morgante, Mortlock, Moss, Munshi,
  Murphy, Naselsky, Nati, Natoli, Netterfield, N{\o}rgaard-Nielsen, Noviello,
  Novikov, Novikov, Oxborrow, Paci, Pagano, Pajot, Paoletti, Partridge, Pasian,
  Patanchon, Pearson, Perdereau, Perotto, Perrotta, Pettorino, Piacentini,
  Piat, Pierpaoli, Pietrobon, Plaszczynski, Pointecouteau, Polenta, Popa,
  Pratt, Pr{\'e}zeau, Prunet, Puget, Rachen, Rebolo, Reinecke, Remazeilles,
  Renault, Renzi, Ristorcelli, Rocha, Roman, Rosset, Rossetti, Roudier,
  Rubi{\~n}o-Mart{\'\i}n, Rusholme, Sandri, Santos, Savelainen, Savini, Scott,
  Seiffert, Shellard, Spencer, Stolyarov, Stompor, Sudiwala, Sunyaev, Sutton,
  Suur-Uski, Sygnet, Tauber, Terenzi, Toffolatti, Tomasi, Tristram, Tucci,
  Tuovinen, T{\"u}rler, Umana, Valenziano, Valiviita, Tent, Vielva, Villa,
  Wade, Wandelt, Wehus, Weller, White, Yvon, Zacchei, \& Zonca}]{ade2016}
Ade, P. a.~R., Aghanim, N., Arnaud, M., {et~al.} 2016,
  \href{http://dx.doi.org/10.1051/0004-6361/201525833}{\JournalTitle{Astronomy
  \& Astrophysics}, 594, A24}

\bibitem[{Aghanim {et~al.}(2016)Aghanim, Arnaud, Ashdown, Aumont, Baccigalupi,
  Banday, Barreiro, Bartlett, Bartolo, Battaner, Benabed, Beno{\^\i}t,
  Benoit-L{\'e}vy, Bernard, Bersanelli, Bielewicz, Bock, Bonaldi, Bonavera,
  Bond, Borrill, Bouchet, Boulanger, Bucher, Burigana, Butler, Calabrese,
  Cardoso, Catalano, Challinor, Chiang, Christensen, Clements, Colombo, Combet,
  Coulais, Crill, Curto, Cuttaia, Danese, Davies, Davis, de~Bernardis, de~Rosa,
  de~Zotti, Delabrouille, D{\'e}sert, Valentino, Dickinson, Diego, Dolag, Dole,
  Donzelli, Dor{\'e}, Douspis, Ducout, Dunkley, Dupac, Efstathiou, Elsner,
  En{\ss}lin, Eriksen, Fergusson, Finelli, Forni, Frailis, Fraisse, Franceschi,
  Frejsel, Galeotta, Galli, Ganga, Gauthier, Gerbino, Giard, Gjerl{\o}w,
  Gonz{\'a}lez-Nuevo, G{\'o}rski, Gratton, Gregorio, Gruppuso, Gudmundsson,
  Hamann, Hansen, Harrison, Helou, Henrot-Versill{\'e},
  Hern{\'a}ndez-Monteagudo, Herranz, Hildebrandt, Hivon, Holmes, Hornstrup,
  Huffenberger, Hurier, Jaffe, Jones, Juvela, Keih{\"a}nen, Keskitalo, Kiiveri,
  Knoche, Knox, Kunz, Kurki-Suonio, Lagache, L{\"a}hteenm{\"a}ki, Lamarre,
  Lasenby, Lattanzi, Lawrence, Jeune, Leonardi, Lesgourgues, Levrier, Lewis,
  Liguori, Lilje, Lilley, Linden-V{\o}rnle, Lindholm, L{\'o}pez-Caniego,
  Mac{\'\i}as-P{\'e}rez, Maffei, Maggio, Maino, Mandolesi, Mangilli, Maris,
  Martin, Mart{\'\i}nez-Gonz{\'a}lez, Masi, Matarrese, Meinhold, Melchiorri,
  Migliaccio, Millea, Mitra, Miville-Desch{\^e}nes, Moneti, Montier, Morgante,
  Mortlock, Mottet, Munshi, Murphy, Narimani, Naselsky, Nati, Natoli, Noviello,
  Novikov, Novikov, Oxborrow, Paci, Pagano, Pajot, Paoletti, Partridge, Pasian,
  Patanchon, Pearson, Perdereau, Perotto, Pettorino, Piacentini, Piat,
  Pierpaoli, Pietrobon, Plaszczynski, Pointecouteau, Polenta, Ponthieu, Pratt,
  Prunet, Puget, Rachen, Reinecke, Remazeilles, Renault, Renzi, Ristorcelli,
  Rocha, Rossetti, Roudier, {d'Orfeuil}, Rubi{\~n}o-Mart{\'\i}n, Rusholme,
  Salvati, Sandri, Santos, Savelainen, Savini, Scott, Serra, Spencer, Spinelli,
  Stolyarov, Stompor, Sunyaev, Sutton, Suur-Uski, Sygnet, Tauber, Terenzi,
  Toffolatti, Tomasi, Tristram, Trombetti, Tucci, Tuovinen, Umana, Valenziano,
  Valiviita, Tent, Vielva, Villa, Wade, Wandelt, Wehus, Yvon, Zacchei, \&
  Zonca}]{aghanim2016}
Aghanim, N., Arnaud, M., Ashdown, M., {et~al.} 2016,
  \href{http://dx.doi.org/10.1051/0004-6361/201526926}{\JournalTitle{Astronomy
  \& Astrophysics}, 594, A11}

\bibitem[{Alam {et~al.}(2017)Alam, Ata, Bailey, Beutler, Bizyaev, Blazek,
  Bolton, Brownstein, Burden, Chuang, Comparat, Cuesta, Dawson, Eisenstein,
  Escoffier, Gil-Mar{\'\i}n, Grieb, Hand, Ho, Kinemuchi, Kirkby, Kitaura,
  Malanushenko, Malanushenko, Maraston, McBride, Nichol, Olmstead, Oravetz,
  Padmanabhan, Palanque-Delabrouille, Pan, Pellejero-Ibanez, Percival,
  Petitjean, Prada, Price-Whelan, Reid, Rodr{\'\i}guez-Torres, Roe, Ross, Ross,
  Rossi, Rubi{\~n}o-Mart{\'\i}n, Saito, Salazar-Albornoz, Samushia,
  S{\'a}nchez, Satpathy, Schlegel, Schneider, Sc{\'o}ccola, Seo, Sheldon,
  Simmons, Slosar, Strauss, Swanson, Thomas, Tinker, Tojeiro, Maga{\~n}a,
  Vazquez, Verde, Wake, Wang, Weinberg, White, Wood-Vasey, Y{\`e}che, Zehavi,
  Zhai, \& Zhao}]{alam2017}
Alam, S., Ata, M., Bailey, S., {et~al.} 2017,
  \href{http://dx.doi.org/10.1093/mnras/stx721}{\JournalTitle{Monthly Notices
  of the Royal Astronomical Society}, 470, 2617}

\bibitem[{Alsing {et~al.}(2018)Alsing, Wandelt, \& Feeney}]{alsing2018}
Alsing, J., Wandelt, B., \& Feeney, S. 2018, \JournalTitle{arXiv:1801.01497
  [astro-ph]}, \href{http://arxiv.org/abs/1801.01497}{{\sffamily
  arXiv:1801.01497 [astro-ph]}}

\bibitem[{Amendola {et~al.}(2016)Amendola, Appleby, Avgoustidis, Bacon, Baker,
  Baldi, Bartolo, Blanchard, Bonvin, Borgani, Branchini, Burrage, Camera,
  Carbone, Casarini, Cropper, {de Rham}, Dietrich, Di~Porto, Durrer, Ealet,
  Ferreira, Finelli, Garcia-Bellido, Giannantonio, Guzzo, Heavens, Heisenberg,
  Heymans, Hoekstra, Hollenstein, Holmes, Horst, Hwang, Jahnke, Kitching,
  Koivisto, Kunz, La~Vacca, Linder, March, Marra, Martins, Majerotto, Markovic,
  Marsh, Marulli, Massey, Mellier, Montanari, Mota, Nunes, Percival, Pettorino,
  Porciani, Quercellini, Read, Rinaldi, Sapone, Sawicki, Scaramella, Skordis,
  Simpson, Taylor, Thomas, Trotta, Verde, Vernizzi, Vollmer, Wang, Weller, \&
  Zlosnik}]{amendola2016}
Amendola, L., Appleby, S., Avgoustidis, A., {et~al.} 2016,
  \JournalTitle{arXiv:1606.00180 [astro-ph]},
  \href{http://arxiv.org/abs/1606.00180}{{\sffamily arXiv:1606.00180
  [astro-ph]}}

\bibitem[{Arthur \& Vassilvitskii(2007)}]{arthur2007}
Arthur, D., \& Vassilvitskii, S. 2007, in Proceedings of the {{Eighteenth
  Annual ACM}}-{{SIAM Symposium}} on {{Discrete Algorithms}}, SODA '07
  (Philadelphia, PA, USA: {Society for Industrial and Applied Mathematics}),
  1027

\bibitem[{Berlind {et~al.}(2006)Berlind, Frieman, Weinberg, Blanton, Warren,
  Abazajian, Scranton, Hogg, Scoccimarro, Bahcall, Brinkmann, Gott, Kleinman,
  Krzesinski, Lee, Miller, Nitta, Schneider, Tucker, Zehavi, \& {SDSS
  Collaboration}}]{berlind2006}
Berlind, A.~A., Frieman, J., Weinberg, D.~H., {et~al.} 2006,
  \href{http://dx.doi.org/10.1086/508170}{\JournalTitle{The Astrophysical
  Journal Supplement Series}, 167, 1}

\bibitem[{Beutler {et~al.}(2017)Beutler, Seo, Saito, Chuang, Cuesta,
  Eisenstein, Gil-Mar{\'\i}n, Grieb, Hand, Kitaura, Modi, Nichol, Olmstead,
  Percival, Prada, S{\'a}nchez, Rodriguez-Torres, Ross, Ross, Schneider,
  Tinker, Tojeiro, \& Vargas-Maga{\~n}a}]{beutler2017}
Beutler, F., Seo, H.-J., Saito, S., {et~al.} 2017,
  \href{http://dx.doi.org/10.1093/mnras/stw3298}{\JournalTitle{Monthly Notices
  of the Royal Astronomical Society}, 466, 2242}

\bibitem[{Bianchi {et~al.}(2015)Bianchi, Gil-Mar{\'\i}n, Ruggeri, \&
  Percival}]{bianchi2015}
Bianchi, D., Gil-Mar{\'\i}n, H., Ruggeri, R., \& Percival, W.~J. 2015,
  \href{http://dx.doi.org/10.1093/mnrasl/slv090}{\JournalTitle{Monthly Notices
  of the Royal Astronomical Society}, 453, L11}

\bibitem[{Bovy {et~al.}(2011)Bovy, Hogg, \& Roweis}]{bovy2011}
Bovy, J., Hogg, D.~W., \& Roweis, S.~T. 2011,
  \href{http://dx.doi.org/10.1214/10-AOAS439}{\JournalTitle{The Annals of
  Applied Statistics}, 5, 1657}

\bibitem[{Broderick {et~al.}(2011)Broderick, Fish, Doeleman, \&
  Loeb}]{broderick2011}
Broderick, A.~E., Fish, V.~L., Doeleman, S.~S., \& Loeb, A. 2011,
  \href{http://dx.doi.org/10.1088/0004-637X/735/2/110}{\JournalTitle{The
  Astrophysical Journal}, 735, 110}

\bibitem[{Cash(1979)}]{cash1979}
Cash, W. 1979, \href{http://dx.doi.org/10.1086/156922}{\JournalTitle{The
  Astrophysical Journal}, 228, 939}

\bibitem[{Collaboration {et~al.}(2014)Collaboration, Ade, Aghanim,
  Armitage-Caplan, Arnaud, Ashdown, Atrio-Barandela, Aumont, Baccigalupi,
  Banday, Barreiro, Barrena, Bartlett, Battaner, Battye, Benabed, Beno{\^\i}t,
  Benoit-L{\'e}vy, Bernard, Bersanelli, Bielewicz, Bikmaev, Blanchard, Bobin,
  Bock, B{\"o}hringer, Bonaldi, Bond, Borrill, Bouchet, Bourdin, Bridges,
  Brown, Bucher, Burenin, Burigana, Butler, Cardoso, Carvalho, Catalano,
  Challinor, Chamballu, Chary, Chiang, Chiang, Chon, Christensen, Church,
  Clements, Colombi, Colombo, Couchot, Coulais, Crill, Curto, Cuttaia,
  Da~Silva, Dahle, Danese, Davies, Davis, {de Bernardis}, {de Rosa}, {de
  Zotti}, Delabrouille, Delouis, D{\'e}mocl{\`e}s, D{\'e}sert, Dickinson,
  Diego, Dolag, Dole, Donzelli, Dor{\'e}, Douspis, Dupac, Efstathiou,
  En{\ss}lin, Eriksen, Finelli, Flores-Cacho, Forni, Frailis, Franceschi,
  Fromenteau, Galeotta, Ganga, G{\'e}nova-Santos, Giard, Giardino,
  Giraud-H{\'e}raud, Gonz{\'a}lez-Nuevo, G{\'o}rski, Gratton, Gregorio,
  Gruppuso, Hansen, Hanson, Harrison, Henrot-Versill{\'e},
  Hern{\'a}ndez-Monteagudo, Herranz, Hildebrandt, Hivon, Hobson, Holmes,
  Hornstrup, Hovest, Huffenberger, Hurier, Jaffe, Jaffe, Jones, Juvela,
  Keih{\"a}nen, Keskitalo, Khamitov, Kisner, Kneissl, Knoche, Knox, Kunz,
  Kurki-Suonio, Lagache, L{\"a}hteenm{\"a}ki, Lamarre, Lasenby, Laureijs,
  Lawrence, Leahy, Leonardi, Le{\'o}n-Tavares, Lesgourgues, Liddle, Liguori,
  Lilje, Linden-V{\o}rnle, L{\'o}pez-Caniego, Lubin, Mac{\'\i}as-P{\'e}rez,
  Maffei, Maino, Mandolesi, Marcos-Caballero, Maris, Marshall, Martin,
  Mart{\'\i}nez-Gonz{\'a}lez, Masi, Matarrese, Matthai, Mazzotta, Meinhold,
  Melchiorri, Melin, Mendes, Mennella, Migliaccio, Mitra,
  Miville-Desch{\^e}nes, Moneti, Montier, Morgante, Mortlock, Moss, Munshi,
  Naselsky, Nati, Natoli, Netterfield, N{\o}rgaard-Nielsen, Noviello, Novikov,
  Novikov, Osborne, Oxborrow, Paci, Pagano, Pajot, Paoletti, Partridge, Pasian,
  Patanchon, Perdereau, Perotto, Perrotta, Piacentini, Piat, Pierpaoli,
  Pietrobon, Plaszczynski, Pointecouteau, Polenta, Ponthieu, Popa, Poutanen,
  Pratt, Pr{\'e}zeau, Prunet, Puget, Rachen, Rebolo, Reinecke, Remazeilles,
  Renault, Ricciardi, Riller, Ristorcelli, Rocha, Roman, Rosset, Roudier,
  Rowan-Robinson, Rubi{\~n}o-Mart{\'\i}n, Rusholme, Sandri, Santos, Savini,
  Scott, Seiffert, Shellard, Spencer, Starck, Stolyarov, Stompor, Sudiwala,
  Sunyaev, Sureau, Sutton, Suur-Uski, Sygnet, Tauber, Tavagnacco, Terenzi,
  Toffolatti, Tomasi, Tristram, Tucci, Tuovinen, T{\"u}rler, Umana, Valenziano,
  Valiviita, Van~Tent, Vielva, Villa, Vittorio, Wade, Wandelt, Weller, White,
  White, Yvon, Zacchei, \& Zonca}]{planckcollaboration2014}
Collaboration, P., Ade, P. A.~R., Aghanim, N., {et~al.} 2014,
  \href{http://dx.doi.org/10.1051/0004-6361/201321521}{\JournalTitle{Astronomy
  \& Astrophysics}, 571, A20}

\bibitem[{Comon(1994)}]{comon1994}
Comon, P. 1994,
  \href{http://dx.doi.org/10.1016/0165-1684(94)90029-9}{\JournalTitle{Signal
  Processing}, 36, 287}

\bibitem[{Crocce {et~al.}(2006)Crocce, Pueblas, \& Scoccimarro}]{crocce2006}
Crocce, M., Pueblas, S., \& Scoccimarro, R. 2006,
  \href{http://dx.doi.org/10.1111/j.1365-2966.2006.11040.x}{\JournalTitle{Monthly
  Notices of the Royal Astronomical Society}, 373, 369}

\bibitem[{Dalal {et~al.}(2008)Dalal, Dor{\'e}, Huterer, \&
  Shirokov}]{dalal2008}
Dalal, N., Dor{\'e}, O., Huterer, D., \& Shirokov, A. 2008,
  \href{http://dx.doi.org/10.1103/PhysRevD.77.123514}{\JournalTitle{Physical
  Review D}, 77}, \href{http://arxiv.org/abs/0710.4560}{{\sffamily
  arXiv:0710.4560}}

\bibitem[{Davis {et~al.}(1985)Davis, Efstathiou, Frenk, \& White}]{davis1985}
Davis, M., Efstathiou, G., Frenk, C.~S., \& White, S. D.~M. 1985,
  \href{http://dx.doi.org/10.1086/163168}{\JournalTitle{The Astrophysical
  Journal}, 292, 371}

\bibitem[{Davison(2008)}]{davison2008}
Davison, A.~C. 2008, Statistical {{Models}} ({{Cambridge Series}} in
  {{Statistical}} and {{Probabilistic Mathematics}}) ({Cambridge University
  Press})

\bibitem[{Dempster {et~al.}(1977)Dempster, Laird, \& Rubin}]{dempster1977}
Dempster, A.~P., Laird, N.~M., \& Rubin, D.~B. 1977, \JournalTitle{Journal of
  the Royal Statistical Society. Series B (Methodological)}, 39, 1

\bibitem[{Efstathiou(2004)}]{efstathiou2004}
Efstathiou, G. 2004,
  \href{http://dx.doi.org/10.1111/j.1365-2966.2004.07530.x}{\JournalTitle{Monthly
  Notices of the Royal Astronomical Society}, 349, 603}

\bibitem[{Efstathiou(2006)}]{efstathiou2006}
---. 2006,
  \href{http://dx.doi.org/10.1111/j.1365-2966.2006.10486.x}{\JournalTitle{Monthly
  Notices of the Royal Astronomical Society}, 370, 343}

\bibitem[{Eifler {et~al.}(2009)Eifler, Schneider, \& Hartlap}]{eifler2009}
Eifler, T., Schneider, P., \& Hartlap, J. 2009,
  \href{http://dx.doi.org/10.1051/0004-6361/200811276}{\JournalTitle{Astronomy
  and Astrophysics}, 502, 721}

\bibitem[{Eisenstein \& Zaldarriaga(2001)}]{eisenstein2001}
Eisenstein, D.~J., \& Zaldarriaga, M. 2001,
  \href{http://dx.doi.org/10.1086/318226}{\JournalTitle{The Astrophysical
  Journal}, 546, 2}

\bibitem[{Feigelson \& Babu(2012)}]{feigelson2012}
Feigelson, E.~D., \& Babu, G.~J. 2012, Modern {{Statistical Methods}} for
  {{Astronomy}}

\bibitem[{Fraley \& Raftery(1998)}]{fraley1998}
Fraley, C., \& Raftery, A.~E. 1998,
  \href{http://dx.doi.org/10.1093/comjnl/41.8.578}{\JournalTitle{The Computer
  Journal}, 41, 578}

\bibitem[{Gardner {et~al.}(2007)Gardner, Connolly, \& McBride}]{gardner2007}
Gardner, J.~P., Connolly, A., \& McBride, C. 2007, in Astronomical {{Data
  Analysis Software}} and {{Systems XVI}}, Vol. 376, 69

\bibitem[{Gazta{\~n}aga \& Scoccimarro(2005)}]{gaztanaga2005}
Gazta{\~n}aga, E., \& Scoccimarro, R. 2005,
  \href{http://dx.doi.org/10.1111/j.1365-2966.2005.09234.x}{\JournalTitle{Monthly
  Notices of the Royal Astronomical Society}, 361, 824}

\bibitem[{Giannantonio {et~al.}(2012)Giannantonio, Porciani, Carron, Amara, \&
  Pillepich}]{giannantonio2012}
Giannantonio, T., Porciani, C., Carron, J., Amara, A., \& Pillepich, A. 2012,
  \href{http://dx.doi.org/10.1111/j.1365-2966.2012.20604.x}{\JournalTitle{Monthly
  Notices of the Royal Astronomical Society}, 422, 2854}

\bibitem[{Giannantonio {et~al.}(2014)Giannantonio, Ross, Percival, Crittenden,
  Bacher, Kilbinger, Nichol, \& Weller}]{giannantonio2014}
Giannantonio, T., Ross, A.~J., Percival, W.~J., {et~al.} 2014,
  \href{http://dx.doi.org/10.1103/PhysRevD.89.023511}{\JournalTitle{Physical
  Review D}, 89}, \href{http://arxiv.org/abs/1303.1349}{{\sffamily
  arXiv:1303.1349}}

\bibitem[{Grieb {et~al.}(2017)Grieb, S{\'a}nchez, Salazar-Albornoz,
  Scoccimarro, Crocce, Dalla~Vecchia, Montesano, Gil-Mar{\'\i}n, Ross, Beutler,
  Rodr{\'\i}guez-Torres, Chuang, Prada, Kitaura, Cuesta, Eisenstein, Percival,
  Vargas-Maga{\~n}a, Tinker, Tojeiro, Brownstein, Maraston, Nichol, Olmstead,
  Samushia, Seo, Streblyanska, \& Zhao}]{grieb2017}
Grieb, J.~N., S{\'a}nchez, A.~G., Salazar-Albornoz, S., {et~al.} 2017,
  \href{http://dx.doi.org/10.1093/mnras/stw3384}{\JournalTitle{Monthly Notices
  of the Royal Astronomical Society}, 467, 2085}

\bibitem[{Guo {et~al.}(2012)Guo, Zehavi, \& Zheng}]{guo2012}
Guo, H., Zehavi, I., \& Zheng, Z. 2012,
  \href{http://dx.doi.org/10.1088/0004-637X/756/2/127}{\JournalTitle{The
  Astrophysical Journal}, 756, 127}

\bibitem[{Hahn {et~al.}(2017{\natexlab{a}})Hahn, Scoccimarro, Blanton, Tinker,
  \& Rodr{\'\i}guez-Torres}]{hahn2017c}
Hahn, C., Scoccimarro, R., Blanton, M.~R., Tinker, J.~L., \&
  Rodr{\'\i}guez-Torres, S.~A. 2017{\natexlab{a}},
  \href{http://dx.doi.org/10.1093/mnras/stx185}{\JournalTitle{Monthly Notices
  of the Royal Astronomical Society}, 467, 1940}

\bibitem[{Hahn {et~al.}(2017{\natexlab{b}})Hahn, Vakili, Walsh, Hearin, Hogg,
  \& Campbell}]{hahn2017b}
Hahn, C., Vakili, M., Walsh, K., {et~al.} 2017{\natexlab{b}},
  \href{http://dx.doi.org/10.1093/mnras/stx894}{\JournalTitle{Monthly Notices
  of the Royal Astronomical Society}, 469, 2791}

\bibitem[{Hand {et~al.}(2017{\natexlab{a}})Hand, Feng, Beutler, Li, Modi,
  Seljak, \& Slepian}]{hand2017b}
Hand, N., Feng, Y., Beutler, F., {et~al.} 2017{\natexlab{a}}

\bibitem[{Hand {et~al.}(2017{\natexlab{b}})Hand, Li, Slepian, \&
  Seljak}]{hand2017a}
Hand, N., Li, Y., Slepian, Z., \& Seljak, U. 2017{\natexlab{b}},
  \href{http://dx.doi.org/10.1088/1475-7516/2017/07/002}{\JournalTitle{Journal
  of Cosmology and Astro-Particle Physics}, 07, 002}

\bibitem[{Hartlap {et~al.}(2009)Hartlap, Schrabback, Simon, \&
  Schneider}]{hartlap2009}
Hartlap, J., Schrabback, T., Simon, P., \& Schneider, P. 2009,
  \href{http://dx.doi.org/10.1051/0004-6361/200911697}{\JournalTitle{Astronomy
  and Astrophysics}, 504, 689}

\bibitem[{Hastie {et~al.}(2009)Hastie, Tibshirani, \& Friedman}]{9780387848587}
Hastie, T., Tibshirani, R., \& Friedman, J. 2009, The {{Elements}} of
  {{Statistical Learning}}: {{Data Mining}}, {{Inference}}, and {{Prediction}},
  {{Second Edition}} ({{Springer Series}} in {{Statistics}}) ({Springer})

\bibitem[{H{\'e}rault \& Ans(1984)}]{herault1984}
H{\'e}rault, J., \& Ans, B. 1984, \JournalTitle{Comptes Rendus de
  l'Acad{\'e}mie des Sciences Paris, S{\'e}rie III, Life Sciences}, 299, 525

\bibitem[{Hogg {et~al.}(2010)Hogg, Bovy, \& Lang}]{hogg2010}
Hogg, D.~W., Bovy, J., \& Lang, D. 2010, \JournalTitle{ArXiv e-prints}, 1008,
  arXiv:1008.4686

\bibitem[{Hu \& White(2001)}]{hu2001}
Hu, W., \& White, M. 2001,
  \href{http://dx.doi.org/10.1086/321380}{\JournalTitle{The Astrophysical
  Journal}, 554, 67}

\bibitem[{Hyv{\"a}rinen(1998)}]{hyvarinen1998}
Hyv{\"a}rinen, A. 1998, in Advances in {{Neural Information Processing
  Systems}} 10, ed. M.~I. Jordan, M.~J. Kearns, \& S.~A. Solla ({MIT Press}),
  273

\bibitem[{Hyvarinen(1999)}]{hyvarinen1999}
Hyvarinen, A. 1999,
  \href{http://dx.doi.org/10.1109/72.761722}{\JournalTitle{IEEE Transactions on
  Neural Networks}, 10, 626}

\bibitem[{Hyvarinen(2001)}]{hyvarinen2001independent}
---. 2001, Independent Component Analysis (New York: {J. Wiley})

\bibitem[{Hyv{\"a}rinen \& Oja(2000)}]{hyvarinen2000}
Hyv{\"a}rinen, A., \& Oja, E. 2000, \JournalTitle{Neural Networks: The Official
  Journal of the International Neural Network Society}, 13, 411

\bibitem[{Kacprzak {et~al.}(2017)Kacprzak, Herbel, Amara, \&
  R{\'e}fr{\'e}gier}]{kacprzak2017}
Kacprzak, T., Herbel, J., Amara, A., \& R{\'e}fr{\'e}gier, A. 2017,
  \JournalTitle{arXiv:1707.07498 [astro-ph]},
  \href{http://arxiv.org/abs/1707.07498}{{\sffamily arXiv:1707.07498
  [astro-ph]}}

\bibitem[{Kazin {et~al.}(2014)Kazin, Koda, Blake, Padmanabhan, Brough, Colless,
  Contreras, Couch, Croom, Croton, Davis, Drinkwater, Forster, Gilbank,
  Gladders, Glazebrook, Jelliffe, Jurek, Li, Madore, Martin, Pimbblet, Poole,
  Pracy, Sharp, Wisnioski, Woods, Wyder, \& Yee}]{kazin2014}
Kazin, E.~A., Koda, J., Blake, C., {et~al.} 2014,
  \href{http://dx.doi.org/10.1093/mnras/stu778}{\JournalTitle{Monthly Notices
  of the Royal Astronomical Society}, 441, 3524}

\bibitem[{Kitaura {et~al.}(2015)Kitaura, Gil-Mar{\'\i}n, Sc{\'o}ccola, Chuang,
  M{\"u}ller, Yepes, \& Prada}]{kitaura2015}
Kitaura, F.-S., Gil-Mar{\'\i}n, H., Sc{\'o}ccola, C.~G., {et~al.} 2015,
  \href{http://dx.doi.org/10.1093/mnras/stv645}{\JournalTitle{Monthly Notices
  of the Royal Astronomical Society}, 450, 1836}

\bibitem[{Kitaura \& He{\ss}(2013)}]{kitaura2013}
Kitaura, F.-S., \& He{\ss}, S. 2013,
  \href{http://dx.doi.org/10.1093/mnrasl/slt101}{\JournalTitle{Monthly Notices
  of the Royal Astronomical Society}, 435, L78}

\bibitem[{Kitaura {et~al.}(2014)Kitaura, Yepes, \& Prada}]{kitaura2014}
Kitaura, F.-S., Yepes, G., \& Prada, F. 2014,
  \href{http://dx.doi.org/10.1093/mnrasl/slt172}{\JournalTitle{Monthly Notices
  of the Royal Astronomical Society}, 439, L21}

\bibitem[{Kitaura {et~al.}(2016)Kitaura, Rodr{\'\i}guez-Torres, Chuang, Zhao,
  Prada, Gil-Mar{\'\i}n, Guo, Yepes, Klypin, Sc{\'o}ccola, Tinker, McBride,
  Reid, S{\'a}nchez, Salazar-Albornoz, Grieb, Vargas-Magana, Cuesta, Neyrinck,
  Beutler, Comparat, Percival, \& Ross}]{kitaura2016}
Kitaura, F.-S., Rodr{\'\i}guez-Torres, S., Chuang, C.-H., {et~al.} 2016,
  \href{http://dx.doi.org/10.1093/mnras/stv2826}{\JournalTitle{Monthly Notices
  of the Royal Astronomical Society}, 456, 4156}

\bibitem[{Klypin {et~al.}(2016)Klypin, Yepes, Gottl{\"o}ber, Prada, \&
  He{\ss}}]{klypin2016}
Klypin, A., Yepes, G., Gottl{\"o}ber, S., Prada, F., \& He{\ss}, S. 2016,
  \href{http://dx.doi.org/10.1093/mnras/stw248}{\JournalTitle{Monthly Notices
  of the Royal Astronomical Society}, 457, 4340}

\bibitem[{Krishnamurthy {et~al.}(2014)Krishnamurthy, Kandasamy, Poczos, \&
  Wasserman}]{krishnamurthy2014}
Krishnamurthy, A., Kandasamy, K., Poczos, B., \& Wasserman, L. 2014,
  \JournalTitle{arXiv:1402.2966 [math, stat]},
  \href{http://arxiv.org/abs/1402.2966}{{\sffamily arXiv:1402.2966 [math,
  stat]}}

\bibitem[{Kuhn \& Feigelson(2017)}]{kuhn2017}
Kuhn, M.~A., \& Feigelson, E.~D. 2017, \JournalTitle{arXiv:1711.11101
  [astro-ph, stat]}, \href{http://arxiv.org/abs/1711.11101}{{\sffamily
  arXiv:1711.11101 [astro-ph, stat]}}

\bibitem[{Lee {et~al.}(2012)Lee, Guillemot, Yue, Kramer, \& Champion}]{lee2012}
Lee, K.~J., Guillemot, L., Yue, Y.~L., Kramer, M., \& Champion, D.~J. 2012,
  \href{http://dx.doi.org/10.1111/j.1365-2966.2012.21413.x}{\JournalTitle{Monthly
  Notices of the Royal Astronomical Society}, 424, 2832}

\bibitem[{Leroux(1992)}]{leroux1992}
Leroux, B.~G. 1992,
  \href{http://dx.doi.org/10.1214/aos/1176348772}{\JournalTitle{The Annals of
  Statistics}, 20, 1350}

\bibitem[{Levi {et~al.}(2013)Levi, Bebek, Beers, Blum, Cahn, Eisenstein,
  Flaugher, Honscheid, Kron, Lahav, McDonald, Roe, Schlegel, \&
  {collaboration}}]{levi2013}
Levi, M., Bebek, C., Beers, T., {et~al.} 2013, \JournalTitle{arXiv:1308.0847
  [astro-ph]}, \href{http://arxiv.org/abs/1308.0847}{{\sffamily arXiv:1308.0847
  [astro-ph]}}

\bibitem[{Liddle(2007)}]{liddle2007}
Liddle, A.~R. 2007,
  \href{http://dx.doi.org/10.1111/j.1745-3933.2007.00306.x}{\JournalTitle{Monthly
  Notices of the Royal Astronomical Society}, 377, L74}

\bibitem[{Lloyd(1982)}]{lloyd1982}
Lloyd, S. 1982,
  \href{http://dx.doi.org/10.1109/TIT.1982.1056489}{\JournalTitle{IEEE
  Transactions on Information Theory}, 28, 129}

\bibitem[{McBride {et~al.}(2009)McBride, Berlind, Scoccimarro, Wechsler, Busha,
  Gardner, \& {van den Bosch}}]{mcbride2009}
McBride, C., Berlind, A., Scoccimarro, R., {et~al.} 2009, in Bulletin of the
  {{American Astronomical Society}}, Vol. 213, 425.06

\bibitem[{McLachlan \& Peel(2000)}]{9780471006268}
McLachlan, G., \& Peel, D. 2000, Finite {{Mixture Models}}
  ({Wiley-Interscience})

\bibitem[{Mohammed {et~al.}(2017)Mohammed, Seljak, \& Vlah}]{mohammed2017}
Mohammed, I., Seljak, U., \& Vlah, Z. 2017,
  \href{http://dx.doi.org/10.1093/mnras/stw3196}{\JournalTitle{Monthly Notices
  of the Royal Astronomical Society}, 466, 780}

\bibitem[{Morrison \& Schneider(2013)}]{morrison2013}
Morrison, C.~B., \& Schneider, M.~D. 2013,
  \href{http://dx.doi.org/10.1088/1475-7516/2013/11/009}{\JournalTitle{Journal
  of Cosmology and Astro-Particle Physics}, 11, 009}

\bibitem[{Neal \& Hinton(1998)}]{neal1998}
Neal, R.~M., \& Hinton, G.~E. 1998,
  \href{http://dx.doi.org/10.1007/978-94-011-5014-9_12}{in Learning in
  {{Graphical Models}}, NATO ASI Series} ({Springer, Dordrecht}), 355

\bibitem[{Norberg {et~al.}(2009)Norberg, Baugh, Gazta{\~n}aga, \&
  Croton}]{norberg2009}
Norberg, P., Baugh, C.~M., Gazta{\~n}aga, E., \& Croton, D.~J. 2009,
  \href{http://dx.doi.org/10.1111/j.1365-2966.2009.14389.x}{\JournalTitle{Monthly
  Notices of the Royal Astronomical Society}, 396, 19}

\bibitem[{Ntampaka {et~al.}(2015)Ntampaka, Trac, Sutherland, Battaglia, Poczos,
  \& Schneider}]{ntampaka2015}
Ntampaka, M., Trac, H., Sutherland, D.~J., {et~al.} 2015,
  \href{http://dx.doi.org/10.1088/0004-637X/803/2/50}{\JournalTitle{The
  Astrophysical Journal}, 803, 50}

\bibitem[{Ntampaka {et~al.}(2016)Ntampaka, Trac, Sutherland, Fromenteau,
  Poczos, \& Schneider}]{ntampaka2016}
---. 2016,
  \href{http://dx.doi.org/10.3847/0004-637X/831/2/135}{\JournalTitle{The
  Astrophysical Journal}, 831, 135}

\bibitem[{O'Connell {et~al.}(2016)O'Connell, Eisenstein, Vargas, Ho, \&
  Padmanabhan}]{oconnell2016}
O'Connell, R., Eisenstein, D., Vargas, M., Ho, S., \& Padmanabhan, N. 2016,
  \href{http://dx.doi.org/10.1093/mnras/stw1821}{\JournalTitle{Monthly Notices
  of the Royal Astronomical Society}, 462, 2681}

\bibitem[{Parkinson {et~al.}(2012)Parkinson, Riemer-S{\o}rensen, Blake, Poole,
  Davis, Brough, Colless, Contreras, Couch, Croom, Croton, Drinkwater, Forster,
  Gilbank, Gladders, Glazebrook, Jelliffe, Jurek, Li, Madore, Martin, Pimbblet,
  Pracy, Sharp, Wisnioski, Woods, Wyder, \& Yee}]{parkinson2012}
Parkinson, D., Riemer-S{\o}rensen, S., Blake, C., {et~al.} 2012,
  \href{http://dx.doi.org/10.1103/PhysRevD.86.103518}{\JournalTitle{Physical
  Review D}, 86, 103518}

\bibitem[{Pinol {et~al.}(2017)Pinol, Cahn, Hand, Seljak, \& White}]{pinol2017}
Pinol, L., Cahn, R.~N., Hand, N., Seljak, U., \& White, M. 2017,
  \href{http://dx.doi.org/10.1088/1475-7516/2017/04/008}{\JournalTitle{Journal
  of Cosmology and Astroparticle Physics}, 2017, 008}

\bibitem[{P{\'o}czos {et~al.}(2011)P{\'o}czos, Szab{\'o}, \&
  Schneider}]{poczos2011}
P{\'o}czos, B., Szab{\'o}, Z., \& Schneider, J. 2011, in 2011 19th {{European
  Signal Processing Conference}}, 1718

\bibitem[{Poczos {et~al.}(2012)Poczos, Xiong, \& Schneider}]{poczos2012}
Poczos, B., Xiong, L., \& Schneider, J. 2012, \JournalTitle{arXiv:1202.3758
  [cs, stat]}, \href{http://arxiv.org/abs/1202.3758}{{\sffamily arXiv:1202.3758
  [cs, stat]}}

\bibitem[{P{\'o}czos {et~al.}(2012)P{\'o}czos, Xiong, Sutherland, \&
  Schneider}]{poczos2012a}
P{\'o}czos, B., Xiong, L., Sutherland, D.~J., \& Schneider, J. 2012,
  \href{http://dx.doi.org/10.1109/CVPR.2012.6248028}{in 2012 {{IEEE
  Conference}} on {{Computer Vision}} and {{Pattern Recognition}}}, 2989

\bibitem[{Press {et~al.}(1992)Press, Teukolsky, Vetterling, \&
  Flannery}]{Press:1992:NRC:148286}
Press, W.~H., Teukolsky, S.~A., Vetterling, W.~T., \& Flannery, B.~P. 1992,
  Numerical {{Recipes}} in {{C}} ({{2Nd Ed}}.): {{The Art}} of {{Scientific
  Computing}} (New York, NY, USA: {Cambridge University Press})

\bibitem[{Ravanbakhsh {et~al.}(2017)Ravanbakhsh, Lanusse, Mandelbaum,
  Schneider, \& Poczos}]{ravanbakhsh2017a}
Ravanbakhsh, S., Lanusse, F., Mandelbaum, R., Schneider, J., \& Poczos, B.
  2017, in Thirty-{{First AAAI Conference}} on {{Artificial Intelligence}}

\bibitem[{Rodr{\'\i}guez-Torres {et~al.}(2016)Rodr{\'\i}guez-Torres, Chuang,
  Prada, Guo, Klypin, Behroozi, Hahn, Comparat, Yepes, Montero-Dorta,
  Brownstein, Maraston, McBride, Tinker, Gottl{\"o}ber, Favole, Shu, Kitaura,
  Bolton, Scoccimarro, Samushia, Schlegel, Schneider, \&
  Thomas}]{rodriguez-torres2016}
Rodr{\'\i}guez-Torres, S.~A., Chuang, C.-H., Prada, F., {et~al.} 2016,
  \href{http://dx.doi.org/10.1093/mnras/stw1014}{\JournalTitle{Monthly Notices
  of the Royal Astronomical Society}, 460, 1173}

\bibitem[{Roeder \& Wasserman(1997)}]{roeder1997}
Roeder, K., \& Wasserman, L. 1997,
  \href{http://dx.doi.org/10.1080/01621459.1997.10474044}{\JournalTitle{Journal
  of the American Statistical Association}, 92, 894}

\bibitem[{Ross {et~al.}(2013)Ross, Percival, Carnero, Zhao, Manera, Raccanelli,
  Aubourg, Bizyaev, Brewington, Brinkmann, Brownstein, Cuesta, {da Costa},
  Eisenstein, Ebelke, Guo, Hamilton, Maga{\~n}a, Malanushenko, Malanushenko,
  Maraston, Montesano, Nichol, Oravetz, Pan, Prada, S{\'a}nchez, Samushia,
  Schlegel, Schneider, Seo, Sheldon, Simmons, Snedden, Swanson, Thomas, Tinker,
  Tojeiro, \& Zehavi}]{ross2013}
Ross, A.~J., Percival, W.~J., Carnero, A., {et~al.} 2013,
  \href{http://dx.doi.org/10.1093/mnras/sts094}{\JournalTitle{Monthly Notices
  of the Royal Astronomical Society}, 428, 1116}

\bibitem[{Ross {et~al.}(2017)Ross, Beutler, Chuang, Pellejero-Ibanez, Seo,
  Vargas-Maga{\~n}a, Cuesta, Percival, Burden, S{\'a}nchez, Grieb, Reid,
  Brownstein, Dawson, Eisenstein, Ho, Kitaura, Nichol, Olmstead, Prada,
  Rodr{\'\i}guez-Torres, Saito, Salazar-Albornoz, Schneider, Thomas, Tinker,
  Tojeiro, Wang, White, \& Zhao}]{ross2017}
Ross, A.~J., Beutler, F., Chuang, C.-H., {et~al.} 2017,
  \href{http://dx.doi.org/10.1093/mnras/stw2372}{\JournalTitle{Monthly Notices
  of the Royal Astronomical Society}, 464, 1168}

\bibitem[{Schwarz(1978)}]{schwarz1978}
Schwarz, G. 1978,
  \href{http://dx.doi.org/10.1214/aos/1176344136}{\JournalTitle{The Annals of
  Statistics}, 6, 461}

\bibitem[{Scoccimarro(1998)}]{scoccimarro1998}
Scoccimarro, R. 1998,
  \href{http://dx.doi.org/10.1046/j.1365-8711.1998.01845.x}{\JournalTitle{Monthly
  Notices of the Royal Astronomical Society}, 299, 1097}

\bibitem[{Scoccimarro(2000)}]{scoccimarro2000}
---. 2000, \href{http://dx.doi.org/10.1086/317248}{\JournalTitle{The
  Astrophysical Journal}, 544, 597}

\bibitem[{Scoccimarro(2015)}]{scoccimarro2015}
---. 2015,
  \href{http://dx.doi.org/10.1103/PhysRevD.92.083532}{\JournalTitle{Physical
  Review D}, 92}, \href{http://arxiv.org/abs/1506.02729}{{\sffamily
  arXiv:1506.02729}}

\bibitem[{Scoccimarro {et~al.}(1999)Scoccimarro, Couchman, \&
  Frieman}]{scoccimarro1999}
Scoccimarro, R., Couchman, H. M.~P., \& Frieman, J.~A. 1999,
  \href{http://dx.doi.org/10.1086/307220}{\JournalTitle{The Astrophysical
  Journal}, 517, 531}

\bibitem[{Scott(1992)}]{scott1992}
Scott, D.~W. 1992, Multivariate {{Density Estimation}}: {{Theory}},
  {{Practice}}, and {{Visualization}} ({Wiley})

\bibitem[{Sellentin \& Heavens(2016)}]{sellentin2016}
Sellentin, E., \& Heavens, A.~F. 2016,
  \href{http://dx.doi.org/10.1093/mnrasl/slv190}{\JournalTitle{Monthly Notices
  of the Royal Astronomical Society}, 456, L132}

\bibitem[{Sellentin \& Heavens(2017)}]{sellentin2017}
---. 2017, \JournalTitle{arXiv:1707.04488 [astro-ph]},
  \href{http://arxiv.org/abs/1707.04488}{{\sffamily arXiv:1707.04488
  [astro-ph]}}

\bibitem[{Sinha {et~al.}(2017)Sinha, Berlind, McBride, Scoccimarro, Piscionere,
  \& Wibking}]{sinha2017}
Sinha, M., Berlind, A.~A., McBride, C.~K., {et~al.} 2017,
  \JournalTitle{arXiv:1708.04892 [astro-ph]},
  \href{http://arxiv.org/abs/1708.04892}{{\sffamily arXiv:1708.04892
  [astro-ph]}}

\bibitem[{Slepian {et~al.}(2017)Slepian, Eisenstein, Brownstein, Chuang,
  Gil-Mar{\'\i}n, Ho, Kitaura, Percival, Ross, Rossi, Seo, Slosar, \&
  Vargas-Maga{\~n}a}]{slepian2017}
Slepian, Z., Eisenstein, D.~J., Brownstein, J.~R., {et~al.} 2017,
  \href{http://dx.doi.org/10.1093/mnras/stx488}{\JournalTitle{Monthly Notices
  of the Royal Astronomical Society}, 469, 1738}

\bibitem[{Slosar {et~al.}(2008)Slosar, Hirata, Seljak, Ho, \&
  Padmanabhan}]{slosar2008}
Slosar, A., Hirata, C., Seljak, U., Ho, S., \& Padmanabhan, N. 2008,
  \href{http://dx.doi.org/10.1088/1475-7516/2008/08/031}{\JournalTitle{Journal
  of Cosmology and Astroparticle Physics}, 2008, 031}

\bibitem[{Springel(2005)}]{springel2005}
Springel, V. 2005,
  \href{http://dx.doi.org/10.1111/j.1365-2966.2005.09655.x}{\JournalTitle{Monthly
  Notices of the Royal Astronomical Society}, 364, 1105}

\bibitem[{Steele \& Raftery(2010)}]{steele2010performance}
Steele, R.~J., \& Raftery, A.~E. 2010

\bibitem[{Sutherland {et~al.}(2012)Sutherland, Xiong, P{\'o}czos, \&
  Schneider}]{sutherland2012}
Sutherland, D.~J., Xiong, L., P{\'o}czos, B., \& Schneider, J. 2012,
  \JournalTitle{arXiv:1202.0302 [cs, stat]},
  \href{http://arxiv.org/abs/1202.0302}{{\sffamily arXiv:1202.0302 [cs, stat]}}

\bibitem[{Taylor {et~al.}(2015)Taylor, Hopkins, Baldry, Bland-Hawthorn, Brown,
  Colless, Driver, Norberg, Robotham, Alpaslan, Brough, Cluver, Gunawardhana,
  Kelvin, Liske, Conselice, Croom, Foster, Jarrett, Lara-Lopez, \&
  Loveday}]{taylor2015}
Taylor, E.~N., Hopkins, A.~M., Baldry, I.~K., {et~al.} 2015,
  \href{http://dx.doi.org/10.1093/mnras/stu1900}{\JournalTitle{Monthly Notices
  of the Royal Astronomical Society}, 446, 2144}

\bibitem[{Tinker \& {et al.}(in preparation)}]{tinkerinpreparation}
Tinker, J.~L., \& {et al.} in preparation

\bibitem[{Vakili \& Hahn(2016)}]{vakili2016}
Vakili, M., \& Hahn, C.~H. 2016, \JournalTitle{arXiv:1610.01991 [astro-ph]},
  \href{http://arxiv.org/abs/1610.01991}{{\sffamily arXiv:1610.01991
  [astro-ph]}}

\bibitem[{Vargas-Maga{\~n}a {et~al.}(2014)Vargas-Maga{\~n}a, Ho, Xu,
  S{\'a}nchez, O'Connell, Eisenstein, Cuesta, Percival, Ross, Aubourg,
  Brownstein, Escoffier, Kirkby, Manera, Schneider, Tinker, \&
  Weaver}]{vargas-magana2014}
Vargas-Maga{\~n}a, M., Ho, S., Xu, X., {et~al.} 2014,
  \href{http://dx.doi.org/10.1093/mnras/stu1681}{\JournalTitle{Monthly Notices
  of the Royal Astronomical Society}, 445, 2}

\bibitem[{Wang {et~al.}(2009)Wang, Sanjeev, \& Sergio}]{wang2009}
Wang, Q., Sanjeev, K., \& Sergio, V. 2009, \JournalTitle{IEEE TRANSACTIONS ON
  INFORMATION THEORY}, 55, 2392

\bibitem[{Wasserman(2004)}]{wasserman2004}
Wasserman, L. 2004, All of {{Statistics}}: {{A Concise Course}} in
  {{Statistical Inference}} ({{Springer Texts}} in {{Statistics}}) ({Springer})

\bibitem[{White \& Padmanabhan(2015)}]{white2015}
White, M., \& Padmanabhan, N. 2015,
  \href{http://dx.doi.org/10.1088/1475-7516/2015/12/058}{\JournalTitle{Journal
  of Cosmology and Astro-Particle Physics}, 12, 058}

\bibitem[{Wilkinson {et~al.}(2015)Wilkinson, Maraston, Thomas, Coccato,
  Tojeiro, Cappellari, Belfiore, Bershady, Blanton, Bundy, Cales, Cherinka,
  Drory, Emsellem, Fu, Law, Li, Maiolino, Masters, Tremonti, Wake, Wang,
  Weijmans, Xiao, Yan, Zhang, Bizyaev, Brinkmann, Kinemuchi, Malanushenko,
  Malanushenko, Oravetz, Pan, \& Simmons}]{wilkinson2015}
Wilkinson, D.~M., Maraston, C., Thomas, D., {et~al.} 2015,
  \href{http://dx.doi.org/10.1093/mnras/stv301}{\JournalTitle{Monthly Notices
  of the Royal Astronomical Society}, 449, 328}

\bibitem[{Wu(1983)}]{wu1983}
Wu, C. F.~J. 1983,
  \href{http://dx.doi.org/10.1214/aos/1176346060}{\JournalTitle{The Annals of
  Statistics}, 11, 95}

\bibitem[{Xu {et~al.}(2013)Xu, Ho, Trac, Schneider, Poczos, \&
  Ntampaka}]{xu2013}
Xu, X., Ho, S., Trac, H., {et~al.} 2013,
  \href{http://dx.doi.org/10.1088/0004-637X/772/2/147}{\JournalTitle{The
  Astrophysical Journal}, 772, 147}

\bibitem[{Zhao {et~al.}(2015)Zhao, Kitaura, Chuang, Prada, Yepes, \&
  Tao}]{zhao2015}
Zhao, C., Kitaura, F.-S., Chuang, C.-H., {et~al.} 2015,
  \href{http://dx.doi.org/10.1093/mnras/stv1262}{\JournalTitle{Monthly Notices
  of the Royal Astronomical Society}, 451, 4266}

\bibitem[{Zheng \& Weinberg(2007)}]{zheng2007}
Zheng, Z., \& Weinberg, D.~H. 2007,
  \href{http://dx.doi.org/10.1086/512151}{\JournalTitle{The Astrophysical
  Journal}, 659, 1}

\end{thebibliography}
\end{document}